\documentclass[fleqn,usenatbib]{mnras}

\usepackage{newtxtext,newtxmath}
\usepackage{anyfontsize}
\usepackage{comment}
\usepackage{multirow}
\usepackage{subcaption}
\usepackage[normalem]{ulem}
\usepackage[dvipsnames]{xcolor}

\usepackage[T1]{fontenc}

\DeclareRobustCommand{\VAN}[3]{#2}
\let\VANthebibliography\thebibliography
\def\thebibliography{\DeclareRobustCommand{\VAN}[3]{##3}\VANthebibliography}

\usepackage{graphicx}	
\usepackage{amsmath}	

\usepackage{hyperref}
\hypersetup{
    colorlinks=true,
    linkcolor=Cerulean,
    citecolor=NavyBlue,
    filecolor=magenta,
    urlcolor=Violet,
}
\usepackage{etoolbox}
\makeatletter
\DeclareRobustCommand\sendemail[4]{
\edef\@tempa{mailto:#1?subject=#2&body=#3 }%
\edef\@tempb{\expandafter\html@spaces\@tempa\@empty}%
\href{\@tempb}{#4}}

\catcode\%=11
\def\html@spaces#1 #2{#1
\catcode\%=14
\makeatother

\usepackage{adjustbox}
\usepackage{textalpha}
\usepackage{textgreek}
\usepackage{xargs}
\usepackage{xspace}

\newcommand{\orcid}[2]{\href{http://orcid.org/#2}{#1}}
\newcommand{\orcidsymb}[2]{\href{http://orcid.org/#2}{#1\adjustbox{trim={-.15\width} {0\height} {-.15\width} {0\height},clip}{\includegraphics[height=10pt]{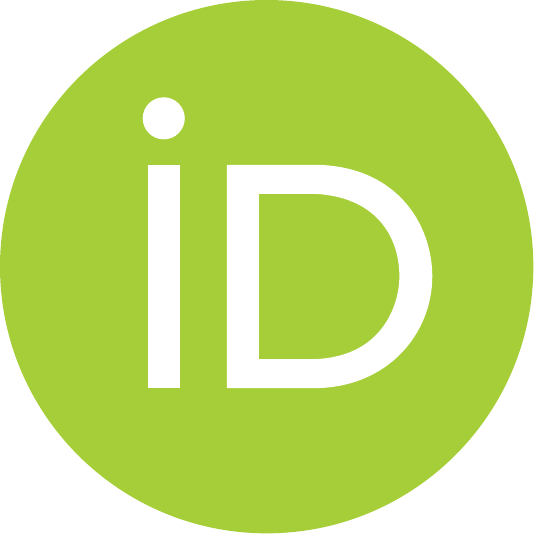}}}}

\newcommand{\citationneeded}{\textcolor{ForestGreen}{$^{\rm citation\;needed}$}}
\let\oldtextdegree\textdegree
\renewcommand{\textdegree}{\oldtextdegree\xspace}
\let\oldtextalpha\textalpha
\renewcommand{\textalpha}{\oldtextalpha\xspace}
\let\oldtextsigma\textsigma
\renewcommand{\textsigma}{\oldtextsigma\xspace}

\newcommand{\kms}{\ensuremath{\mathrm{km\,s^{-1}}}\xspace}
\newcommand{\Msun}{\ensuremath{{\rm M}_\odot}\xspace}
\newcommand{\Zsun}{\ensuremath{{\rm Z}_\odot}\xspace}
\newcommand{\yr}{\ensuremath{{\rm yr}}\xspace}
\newcommand{\Myr}{\ensuremath{{\rm Myr}}\xspace}
\newcommand{\Gyr}{\ensuremath{{\rm Gyr}}\xspace}
\newcommand{\peryr}{\ensuremath{{\rm yr^{-1}}}\xspace}
\newcommand{\Lsun}{\hbox{\,${\rm L}_\odot$}}
\newcommand{\kpc}{\text{kpc}\xspace}
\newcommand{\ZH}{\text{[Z/H]}\xspace}
\newcommand{\MUV}{\ensuremath{M_\mathrm{UV}}\xspace}
\newcommandx{\percm}[1][1=3]{\ensuremath{\mathrm{cm}^{-#1}}\xspace}	

\newcommandx{\lambdar}[2][1=R,2=]{\ensuremath{\lambda_{\rm {#1}}{#2}}\xspace}
\newcommand{\eps}{\ensuremath{\epsilon}\xspace}
\newcommand{\mstar}{\ensuremath{M_\star}\xspace}
\newcommand{\mdyn}{\ensuremath{M_\mathrm{dyn}}\xspace}
\newcommand{\re}{\ensuremath{R_\mathrm{e}}\xspace}
\newcommand{\vstar}{\ensuremath{v_\star}\xspace}
\newcommand{\vnai}{\ensuremath{v_{\NaI}}\xspace}
\newcommand{\sigmastar}{\ensuremath{\sigma_\star}\xspace}
\newcommand{\sigmaestar}{\ensuremath{\sigma_{\star,\mathrm{e}}}\xspace}
\newcommand{\vperc}[1]{\ensuremath{v_{#1}}\xspace}

\newcommand{\vesc}{\ensuremath{v_\mathrm{esc}}\xspace}
\newcommand{\nelec}{\ensuremath{n_\mathrm{e}}\xspace}
\newcommand{\Telec}{\ensuremath{T_\mathrm{e}}\xspace}
\newcommand{\Rout}{\ensuremath{R_\mathrm{out}}\xspace}
\newcommand{\vout}{\ensuremath{v_\mathrm{out}}\xspace}
\newcommandx{\Mout}[2][1=,2=]{\ensuremath{M_{\mathrm{out}{#2}}^{#1}}\xspace}
\newcommandx{\Mdotout}[2][1=,2=]{\ensuremath{\dot{M}_{\mathrm{out}{#2}}^{#1}}\xspace}
\newcommand\sbullet[1][.5]{\mathbin{\vcenter{\hbox{\scalebox{#1}{$\bullet$}}}}}
\newcommand{\mbh}{\ensuremath{M_{\sbullet[0.85]}}\xspace}

\newcommandx{\fluxdcgs}[3][1=-20,2=\times,3=]{\ensuremath{{#2}10^{#1}~\mathrm{erg\,s^{{#3}-{#3}1}\,cm^{{#3}-{#3}2}\,\AA^{{#3}-{#3}1}}}\xspace}
\newcommandx{\fluxcgs}[3][1=-20,2=\times,3=]{\ensuremath{{#2}10^{#1}~\mathrm{erg\,s^{{#3}-{#3}1}\,cm^{{#3}-{#3}2}}}\xspace}
\newcommandx{\powercgs}[2][1=44,2=\times]{\ensuremath{{#2}10^{#1}~\mathrm{erg\,s^{-1}}}\xspace}
\newcommandx{\ergs}{\ensuremath{\mathrm{erg\,s^{-1}}}\xspace}
\newcommand{\AV}{\ensuremath{A_V}\xspace}

\newcommand{\cigale}{{\sc cigale}\xspace}
\newcommand{\jwst}{\textit{JWST}\xspace}
\newcommand{\hst}{\textit{HST}\xspace}
\newcommand{\ppxf}{{\sc ppxf}\xspace}
\newcommand{\prospector}{{\sc prospector}\xspace}
\newcommand{\emcee}{{\sc emcee}\xspace}
\newcommand{\eazy}{{\sc eazy}\xspace}
\newcommand{\cloudy}{{\sc cloudy}\xspace}
\newcommand{\pyneb}{{\sc pyneb}\xspace}
\newcommandx{\mappings}[1][1=]{{\sc mappings{#1}}\xspace}
\newcommand{\galfit}{{\sc galfit}\xspace}
\newcommand{\beagle}{{\sc beagle}\xspace}
\newcommand{\qubespec}{{\sc qubespec}\xspace}
\newcommand{\pysersic}{{\sc pysersic}\xspace}
\newcommand{\mpt}{{\sc mpt}\xspace}

\newcommand{\blackthunder}{BlackTHUNDER\xspace}
\newcommand{\Mdynvalue}{$\Mdyn = 2.0\pm0.5 \times 10^{11}$~\MSun}

\defcitealias{gordon+2003}{G03}
\defcitealias{nakajima+maiolino2022}{NM22}

\usepackage{amsmath}	
\usepackage{textalpha}
\usepackage{textgreek}
\usepackage{xargs}
\usepackage{xspace}

\let\oldAA\AA
\renewcommand{\AA}{\text{\oldAA}\xspace}
\newcommand{\mum}{\text{\textmu m}\xspace}

\newcommand{\Lyalpha}{\text{Ly\,\textalpha}\xspace}
\newcommand{\Halpha}{\text{H\,\textalpha}\xspace}
\newcommand{\Hbeta}{\text{H\,\textbeta}\xspace}
\newcommand{\Hgamma}{\text{H\,\textgamma}\xspace}
\newcommand{\Hdelta}{\text{H\,\textdelta}\xspace}
\newcommand{\Paalpha}{\text{Pa\,\textalpha}\xspace}
\newcommand{\Pabeta}{\text{Pa\,\textbeta}\xspace}
\newcommand{\Pagamma}{\text{Pa\,\textgamma}\xspace}
\newcommand{\Padelta}{\text{Pa\,\textdelta}\xspace}
\newcommand{\Hepsilon}{\text{H\,\textepsilon}\xspace}
\newcommand{\Hzeta}{\text{H\,\textzeta}\xspace}

\newcommandx{\permittedEL}[6][1=O,2=III,3=,4=,5=,6=]{\text{{#1}\,{\sc {#2}}{#3}{#4}{#5}{#6}}\xspace}
\newcommandx{\semiforbiddenEL}[6][1=O,2=III,3=,4=,5=,6=]{\text{{#1}\,{\sc{#2}}]{#3}{#4}{#5}{#6}}\xspace}
\newcommandx{\forbiddenEL}[6][1=O,2=III,3=,4=,5=,6=]{\text{[{#1}\,{\sc{#2}}]{#3}{#4}{#5}{#6}}\xspace}

\newcommand{\HI}{\permittedEL[H][i]}
\newcommand{\HII}{\permittedEL[H][ii]}

\newcommand{\NV}{\permittedEL[N][v]}
\newcommandx{\NVL}[1][1=1243]{\permittedEL[N][v][\textlambda][#1]}
\newcommandx{\NVall}{\permittedEL[N][v][\textlambda][\textlambda][1239,][1243]}

\newcommandx{\CIIall}{\semiforbiddenEL[C][ii][\textlambda][\textlambda][2324--][2329]}

\newcommand{\NIV}{\semiforbiddenEL[N][iv]}
\newcommandx{\NIVL}[1][1=1486]{\semiforbiddenEL[N][iv][\textlambda][#1]}

\newcommand{\CIV}{\permittedEL[C][iv]}
\newcommandx{\CIVL}[1][1=1550]{\permittedEL[C][iv][\textlambda][#1]}
\newcommand{\CIVall}{\permittedEL[C][iv][\textlambda][\textlambda][1548,][1551]}

\newcommand{\HeII}{\permittedEL[He][ii]}
\newcommandx{\HeIIL}[1][1=1640]{\permittedEL[He][ii][\textlambda][#1]}

\newcommand{\semiOIII}{\semiforbiddenEL[O][iii]}
\newcommandx{\semiOIIIL}[1][1=1666]{\semiforbiddenEL[O][iii][\textlambda][#1]}
\newcommand{\semiOIIIall}{\semiforbiddenEL[O][iii][\textlambda][\textlambda][1661,][1666]}

\newcommand{\NIII}{\semiforbiddenEL[N][iii]}
\newcommandx{\NIIIL}[1][1=1750]{\semiforbiddenEL[N][iii][\textlambda][#1]}
\newcommand{\NIIIall}{\semiforbiddenEL[N][iii][\textlambda][\textlambda][1747--][1754]}

\newcommandx{\CIII}{\semiforbiddenEL[C][iii]}
\newcommandx{\CIIIL}[1][1=1909]{\semiforbiddenEL[C][iii][\textlambda][#1]}
\newcommand{\CIIIall}{\semiforbiddenEL[C][iii][\textlambda][\textlambda][1907,][1909]}

\newcommand{\NeIV}{\forbiddenEL[Ne][iv]}
\newcommandx{\NeIVL}[1][1=2424]{\forbiddenEL[Ne][iv][\textlambda][#1]}
\newcommand{\NeIVall}{\forbiddenEL[Ne][iv][\textlambda][\textlambda][2422,][2424]}

\newcommand{\MgII}{\permittedEL[Mg][ii]}
\newcommandx{\MgIIL}[1][1=2803]{\permittedEL[Mg][ii][\textlambda][#1]}
\newcommand{\MgIIall}{\permittedEL[Mg][ii][\textlambda][\textlambda][2796,][2803]}

\newcommand{\NeV}{\forbiddenEL[Ne][v]}
\newcommandx{\NeVL}[1][1=3426]{\forbiddenEL[Ne][v][\textlambda][#1]}
\newcommand{\NeVall}{\forbiddenEL[Ne][v][\textlambda][\textlambda][3346,][3426]}

\newcommand{\OIIperm}{\permittedEL[O][ii]}
\newcommand{\OII}{\forbiddenEL[O][ii]}
\newcommandx{\OIIL}[1][1=3726]{\forbiddenEL[O][ii][\textlambda][#1]}
\newcommandx{\OIIall}[2][1={3726,},2=3729]{\forbiddenEL[O][ii][\textlambda][\textlambda][#1][#2]}

\newcommand{\NeIII}{\forbiddenEL[Ne][iii]}
\newcommandx{\NeIIIL}[1][1=3869]{\forbiddenEL[Ne][iii][\textlambda][#1]}
\newcommand{\NeIIIall}{\forbiddenEL[Ne][iii][\textlambda][\textlambda][3869,][3967]}

\newcommand{\OIII}{\forbiddenEL[O][iii]}
\newcommandx{\OIIIL}[1][1=5007]{\forbiddenEL[O][iii][\textlambda][#1]}
\newcommand{\OIIIall}{\forbiddenEL[O][iii][\textlambda][\textlambda][4959,][5007]}

\newcommandx{\NIL}[1][1=5200]{\forbiddenEL[N][i][\textlambda][#1]}
\newcommand{\NIall}{\forbiddenEL[N][i][\textlambda][\textlambda][5198,][5200]}

\newcommand{\OI}{\forbiddenEL[O][i]}
\newcommandx{\OIL}[1][1=6300]{\forbiddenEL[O][i][\textlambda][#1]}
\newcommand{\OIall}{\forbiddenEL[O][i][\textlambda][\textlambda][6300,][6364]}

\newcommand{\HeI}{\permittedEL[He][i]}
\newcommandx{\HeIL}[1][1=5875]{\permittedEL[He][i][\textlambda][#1]}

\newcommand{\OIperm}{\permittedEL[O][i]}
\newcommand{\OIres}{\permittedEL[O][i]}
\newcommandx{\OIresL}[1][1=8446]{\permittedEL[O][i][\textlambda][#1]}

\newcommand{\NII}{\forbiddenEL[N][ii]}
\newcommandx{\NIIL}[1][1=6583]{\forbiddenEL[N][ii][\textlambda][#1]}
\newcommand{\NIIall}{\forbiddenEL[N][ii][\textlambda][\textlambda][6548,][6583]}

\newcommand{\SII}{\forbiddenEL[S][ii]}
\newcommandx{\SIIL}[1][1=6716]{\forbiddenEL[S][ii][\textlambda][#1]}
\newcommandx{\SIIall}[2][1=6716,2=6731]{\forbiddenEL[S][ii][\textlambda][\textlambda][{#1},][#2]}

\newcommand{\SIII}{\forbiddenEL[S][iii]}
\newcommandx{\SIIIL}[1][1=9532]{\forbiddenEL[S][iii][\textlambda][#1]}
\newcommandx{\SIIIall}[2][1=9069,2=9532]{\forbiddenEL[S][iii][\textlambda][\textlambda][{#1},][#2]}

\newcommandx{\OIIAuL}[1][1=7325]{\forbiddenEL[O][ii][\textlambda][#1]}
\newcommand{\OIIAuall}{\forbiddenEL[O][ii][\textlambda][\textlambda][7319--][7331]}

\newcommandx{\CIIFIRL}{\forbiddenEL[C][ii][\textlambda][158\,\mum]}

\newcommand{\NaI}{\permittedEL[Na][i]}
\newcommandx{\NaIL}[1][1=5890]{\permittedEL[Na][i][\textlambda][#1]}
\newcommand{\NaIall}{\permittedEL[Na][i][\textlambda][\textlambda][5890,][5896]}

\newcommand{\CaII}{\permittedEL[Ca][ii]}
\newcommandx{\CaIIL}[1][1=3934]{\permittedEL[Ca][ii][\textlambda][#1]}
\newcommand{\CaIIall}{\permittedEL[Ca][ii][\textlambda][\textlambda][3934,][3968]}

\newcommandx{\FeIIperm}{\permittedEL[Fe][ii]}
\newcommandx{\FeII}{\forbiddenEL[Fe][ii]}
\newcommandx{\FeVII}{\forbiddenEL[Fe][vii]}
\newcommandx{\FeIIL}[1][1=5159]{\forbiddenEL[Fe][ii][\textlambda][#1]}
\newcommandx{\FeIIall}{\forbiddenEL[Fe][ii][\textlambda][\textlambda][4359,][4414]}

\newcommand{\targetlong}{SDSS~J102530.29+140207.3\xspace}
\newcommand{\target}{J1025\xspace}
\newcommand{\lrdstar}{HD~51585\xspace}
\newcommand{\betauv}{\ensuremath{\beta_\text{UV}}\xspace}
\newcommand{\logU}{\ensuremath{\log\,U}\xspace}
\newcommand{\lEdd}{\ensuremath{\lambda_\text{Edd}}\xspace}

\newcommand{\qc}{\ensuremath{q_{\rm c}}\xspace}
\newcommand{\uc}{\ensuremath{u_{\rm c}}\xspace}
\newcommand{\ql}{\ensuremath{q_{\Halpha}}\xspace}
\newcommand{\ul}{\ensuremath{u_{\Halpha}}\xspace}
\newcommand{\mailmessage}{Dear Francesco and Gabriele,\%0A\%0ASince when do you do spectropolarimetry? When we were kids, there was this cousin who also had a spectropolarimeter in his garage - oh, these were fun times... Nevermind, back to the science question: in the paper, I was really surprised by your claim that...\%0A\%0ARegards,\%0A}

\newcommand\blfootnote[1]{%
  \begingroup
  \renewcommand\thefootnote{}\footnote{#1}%
  \addtocounter{footnote}{-1}%
  \endgroup
}

\newcommand\samethanks[1][\value{footnote}]{\footnotemark[#1]}

\title[Polarisation of a Little Red Dot]{Misaligned or chaotic? A strong break of axial symmetry in the local LRD J1025 revealed with VLT/FORS2 spectropolarimetry}

\author[\sendemail{francesco.deugenio@gmail.com,pezzulli@astro.rug.nl}{Question about your Spectro-polarimetry paper}{\mailmessage}{F. D'Eugenio \& G. Pezzulli}~et al.]{\parbox{\textwidth}{
\orcidsymb{Francesco D'Eugenio}{0000-0003-2388-8172}$^{\hyperlink{aff1}{1},\hyperlink{aff2}{2}}$\thanks{These authors contributed equally to the article.}\thanks{E-mail: \sendemail{francesco.deugenio@gmail.com,pezzulli@astro.rug.nl}{Question about your Spectro-polarimetry paper}{\mailmessage}{francesco.deugenio@gmail.com}},
\orcidsymb{Gabriele Pezzulli}{0000-0003-0736-7879}$^{\hyperlink{aff3}{3}}$\hyperlink{Hfootnote.1}{\ensuremath{^\star}}\thanks{E-mail: \sendemail{francesco.deugenio@gmail.com,pezzulli@astro.rug.nl}{Question about your Spectro-polarimetry paper}{\mailmessage}{pezzulli@astro.rug.nl}},
\orcidsymb{Roberto Maiolino}{0000-0002-4985-3819}$^{\hyperlink{aff1}{1},\hyperlink{aff2}{2},\hyperlink{aff4}{4}}$,
\orcidsymb{Alessandro Marconi}{0000-0002-9889-4238}$^{\hyperlink{aff5}{5},\hyperlink{aff6}{6}}$,
\orcidsymb{Xihan Ji}{0000-0002-1660-9502}$^{\hyperlink{aff1}{1},\hyperlink{aff2}{2}}$,
\orcidsymb{Cristina Ramos Almeida}{0000-0001-8353-649X}$^{\hyperlink{aff7}{7},\hyperlink{aff8}{8}}$,
\orcidsymb{Andrea Ferrara}{0000-0002-9400-7312}$^{\hyperlink{aff9}{9}}$,
\orcidsymb{Piero Madau}{0000-0002-6336-3293}$^{\hyperlink{aff10}{10}}$,
\orcidsymb{Xiaojing Lin}{0000-0001-6052-4234}$^{\hyperlink{aff11}{11},\hyperlink{aff12}{12}}$,
\orcidsymb{Jos\'e A. Acosta-Pulido}{0000-0002-0433-9656}$^{\hyperlink{aff7}{7},\hyperlink{aff8}{8}}$,
\orcidsymb{Fuyan Bian}{0000-0002-1620-0897}$^{\hyperlink{aff13}{13},\hyperlink{aff14}{14}}$,
\orcidsymb{Matilde Brazzini}{0009-0009-6418-7154}$^{\hyperlink{aff15}{15},\hyperlink{aff16}{16},\hyperlink{aff14}{14},\hyperlink{aff1}{1},\hyperlink{aff2}{2}}$,
\orcidsymb{Zheng Cai}{0000-0001-8467-6478}$^{\hyperlink{aff11}{11}}$,
\orcidsymb{Stefano Carniani}{0000-0002-6719-380X}$^{\hyperlink{aff9}{9}}$,
\orcidsymb{Xiaohui Fan}{0000-0002-2662-8803}$^{\hyperlink{aff12}{12}}$,
\orcidsymb{Ignas Juod\v{z}balis}{0009-0003-7423-8660}$^{\hyperlink{aff1}{1},\hyperlink{aff2}{2}}$,
\orcidsymb{Robert G. Pascalau}{0000-0001-9820-5773}$^{\hyperlink{aff1}{1},\hyperlink{aff2}{2}}$,
\orcidsymb{Jan Scholtz}{0000-0001-6010-6809}$^{\hyperlink{aff1}{1},\hyperlink{aff2}{2}}$,
\orcidsymb{Charlotte Simmonds}{0000-0003-4770-7516}$^{\hyperlink{aff17}{17}}$,
\orcidsymb{Fengwu Sun}{0000-0002-4622-6617}$^{\hyperlink{aff18}{18}}$,
and \orcidsymb{Sandro Tacchella}{0000-0002-8224-4505}$^{\hyperlink{aff1}{1},\hyperlink{aff2}{2}}$
}\vspace{0.4cm}
\\
\parbox{\textwidth}{
\hypertarget{aff1}{$^{1}$}Kavli Institute for Cosmology, University of Cambridge, Madingley Road, Cambridge CB3 0HA, UK\\
\hypertarget{aff2}{$^{2}$}Cavendish Laboratory, University of Cambridge, JJ Thomson Avenue, Cambridge CB3 0HE, UK\\
\hypertarget{aff3}{$^{3}$}Kapteyn Astronomical Institute, University of Groningen, Landleven 12, 9747 AD Groningen, The Netherlands\\
\hypertarget{aff4}{$^{4}$}Department of Physics and Astronomy, University College London, Gower Street, London WC1E 6BT, UK\\
\hypertarget{aff5}{$^{5}$}Dipartimento di Fisica e Astronomia, Universit\`a degli Studi di Firenze, Via G. Sansone 1, 50019 Sesto Fiorentino (Firenze), Italy\\
\hypertarget{aff6}{$^{6}$}INAF -- Osservatorio Astrofisico di Arcetri, Largo E. Fermi 5, 50125 Firenze, Italy\\
\hypertarget{aff7}{$^{7}$}Instituto de Astrof\'isica de Canarias, Calle V\'ia L\'actea s/n, E-38205 La Laguna, Tenerife, Spain\\
\hypertarget{aff8}{$^{8}$}Departamento de Astrof\'isica, Universidad de La Laguna, E-38206 La Laguna, Tenerife, Spain\\
\hypertarget{aff9}{$^{9}$}Scuola Normale Superiore, Piazza dei Cavalieri 7, 56126 Pisa, Italy\\
\hypertarget{aff10}{$^{10}$}Department of Astronomy and Astrophysics, University of California, Santa Cruz, 1156 High Street, Santa Cruz, CA 95064, USA\\
\hypertarget{aff11}{$^{11}$}Department of Astronomy, Tsinghua University, Beijing 100084, China\\
\hypertarget{aff12}{$^{12}$}Steward Observatory, University of Arizona, 933 N. Cherry Avenue, Tucson, AZ 85721, USA\\
\hypertarget{aff13}{$^{13}$}European Southern Observatory, Alonso de Córdova 3107, Casilla 19001, Vitacura, Santiago 19, Chile\\
\hypertarget{aff14}{$^{14}$}Chinese Academy of Sciences South America Center for Astronomy, National Astronomical Observatories, CAS, Beijing 100101, People's Republic of China\\
\hypertarget{aff15}{$^{15}$}Dipartimento di Fisica, Sezione di Astronomia, Universit\`a degli Studi di Trieste, Via G.B. Tiepolo 11, 34143 Trieste, Italy\\
\hypertarget{aff16}{$^{16}$}INAF -- Osservatorio Astronomico di Trieste, Via G. B. Tiepolo 11, I-34143 Trieste, Italy\\
\hypertarget{aff17}{$^{17}$}Departamento de Astronom{\' i}a, Universidad de Chile, Camino El Observatorio 1515, Las Condes, Santiago, Chile\\
\hypertarget{aff18}{$^{18}$}Center for Astrophysics $|$ Harvard \& Smithsonian, 60 Garden Street, Cambridge, MA 02138, USA\\
}}

\date{Accepted XXX. Received YYY; in original form ZZZ}
\pubyear{2026}

\begin{document}
\label{firstpage}
\pagerange{\pageref{firstpage}--\pageref{lastpage}}
\maketitle

\begin{abstract}
Little Red Dots (LRDs) are compact active galactic nuclei (AGN) with unusual spectral energy distributions and broad Balmer emission, candidate signposts of rapid black-hole growth. We present VLT/FORS2 optical linear spectropolarimetry of the closest known LRD, \targetlong, at $z=0.1$. In total light, we detect
spatially extended narrow-\Halpha emission, probably tracing the host galaxy. We measure a nearly grey continuum polarisation $p_{\rm cont}=1.53\pm0.04$(rand.)$\pm0.20$(syst.) per cent, while broad \Halpha is less polarised, $p_{\Halpha}=0.58\text{--}0.84$ per cent. We rule out polarisation by Milky Way dust and dilution by an unpolarised line. The polarised continuum with a depolarised line resembles local Seyfert-1 nuclei and favours a single dominant source over multi-source explanations. After Stokes-continuum subtraction, broad \Halpha shows no blue-to-red swing, but there is a significant $(48\pm4)$\textdegree\ continuum-to-line offset in polarisation angle. This implies a break of axial symmetry inside this object, posing interesting geometrical challenges to all existing LRD models and frameworks. We discuss different possible origins of this symmetry break and possible paths to discriminate between them with future observations.
\end{abstract}


\section{Introduction}\label{s.intro}

JWST has uncovered a populous class of compact, optically red sources at $z\gtrsim4$ showing broad Balmer (and often He) emission and a characteristic `V-shaped' UV--optical continuum, dubbed `Little Red Dots' \citep[LRDs;][]{ matthee+2024,greene+2024,kocevski+2025}. Their physical nature is still debated, because several of their properties are atypical for standard unobscured active galactic nuclei (AGN). They are strong in the optical, but weak at wavelengths where AGN often dominate: X-rays \citep{simmonds+2016,yue+2024,ananna+2024,lambrides+2024,pacucci+2024,maiolino+2025}, high-ionization lines \citep{lambrides+2024,zucchi+2025,wang+2025,ji+2026a}, mid-infrared \citetext{MIR; \citealp{akins+2025,setton+2025,barro+2025,ronayne+2025,leung+2024,wang+2025b}, but see also \citealp{delvecchio+2025,brazzini+2026}}, and radio \citep{mazzolari+2025}. They also host black holes that seem to be `overmassive' relative to their host stellar mass and to the expectation set by the local black-hole mass--stellar-mass relation \citep[\mbh--\mstar;][]{harikane+2023,maiolino+2024,greene+2024, taylor+2025,juodzbalis+2026a,jones+2026}.

Noticing that a significant fraction of LRDs are characterised by exponential wings in their broad-line profiles, it has been proposed that the width of the broad Balmer lines arises primarily from electron scattering in dense ionized gas \citep{rusakov+2026}, which would imply substantially narrower intrinsic line widths and hence much smaller \mbh. The electron scattering interpretation is attractive for two reasons. First, because it naturally explains the exponential wing shape in many LRDs \citep{rusakov+2026,torralba+2026b,matthee+2026}; second, by lowering \mbh, it resolves the problem of overmassive black holes \citep{maiolino+2024}\footnote{\citet{maiolino+2024} reports that while LRDs exceed the local \mbh--\mstar relation, they seem to lie on the local \mbh--$\sigma$ relation, under the assumption that the measured velocity dispersion of the ionised gas $\sigma$ is a faithful predictor for the velocity dispersion of the underlying stellar population \citep[see also][]{mcclymont+2026,ortame+2026}. Reconciling the \mbh--\mstar relation may therefore be in tension with the \mbh--$\sigma$ relation.}, while implying a rapid growth (as the inferred Eddington ratio is higher if \mbh is lower \citep{rusakov+2026}), thus revealing an evolutionary phase of intermediate mass and rapid growth that has remained elusive since the discovery of SMBH at high redshift \citep{fan+2001,inayoshi+2020}. On the other hand, for at least one object, the lensed LRD Abell2744-QSO1 \citep{furtak+2024}, there is a direct dynamical \mbh measurement from spatially resolved ionised-gas kinematics. The inferred \mbh for this object is substantially higher than what implied by the electron scattering model, indicating that the origin of the line broadening is not yet settled and warrants further investigation. It should also be noted that exponential Balmer-line wings are not unique to LRDs. Standard type-1 AGN are known to display a range of non-Gaussian broad-line profiles \citep[e.g.,][]{nagao+2006,kollatschny+zetzl2013}, and a large subset of these AGN displays exponential wings (Trefoloni et al., in~prep.), without sharing the other anomalies found in LRDs. In particular, reverberation mapping and even interferometric measurements support interpreting the heavy tails in these standard AGN as due to virial broadening, and in some cases to inner outflows \citep{gravity+2018,gravity+2024,gravity+2026}, not radiative transfer effects. This suggests that a similar interpretation may apply to LRDs too. Motivated by this, a framework has been recently introduced to model the exponential wings of LRDs with emission from a stratified virialised BLR with no broadening from electron scattering \citep{scholtz+2026b,madau+2026}.

Another key open question is the joint role of orientation and dust in LRDs: in the classical unified model of AGN, the inclination of a dusty, obscuring torus has an overwhelming impact on the observed SED \citep[e.g.,][]{urry+padovani1995}. For LRDs, the hot dust component was initially found to be surprisingly faint compared to expectations \citep[e.g.,][]{setton+2025}, although it is detected in several objects, with a luminosity that varies strongly from source to source \citep[e.g.,][]{juodzbalis+2024,degraaff+2025a} and correlates with the continuum spectral properties \citep{perez-gonzalez+2026}. At the same time, distinctive features indicative of dense gas absorption are commonly detected in LRDs \citep[e.g.,][]{inayoshi+maiolino2025,degraaff+2025a,deugenio+2026a}. It remains debated whether the exceptional properties of LRDs require an entirely separate paradigm \citep[e.g.,][]{naidu+2025} or instead at least some properties of the standard unification model can be applied also to LRDs \citep[e.g.,][]{madau+maiolino2026}.

Progress on the LRD puzzle has been boosted by the discovery of local counterparts at $z\simeq0.1\text{--}0.2$, which share the defining spectral features of high-redshift LRDs while permitting high signal-to-noise follow-up \citep{lin+2026a,ji+2026a}. In particular, \targetlong \citetext{`The Egg' in \citealt{lin+2026a}; `Lord of LRDs' in \citealt{ji+2026a}; hereafter: \target} exhibits broad Balmer lines, a `v'-shaped spectral energy distribution (SED), deep Balmer-line absorption, forbidden \FeII emission \citep{lin+2026a,ji+2026a} and extreme X-ray weakness \citep{ji+2026a}. These nearby systems provide a unique laboratory to test existing theories of line broadening, dust obscuration and orientation effects in LRDs.

Polarimetry has historically been a decisive tool for mapping AGN structure and testing unification by orientation. Bypassing any obscuring medium, scattered light can effectively provide an additional, different sightline towards obscured systems. The spectropolarimetric detection of broad permitted lines in polarised flux in NGC~1068 provided direct evidence for a luminous, hidden BLR whose direct view is blocked by optically thick circumnuclear material, while scattered light escapes along less-obscured directions \citep{antonucci+miller1985}. This result became a cornerstone of unified AGN models, in which type~1 and type~2 classifications largely reflect viewing angle relative to an obscuring torus and associated scattering regions \citep{antonucci1993}. Subsequent spectropolarimetric surveys demonstrated that a substantial fraction of Seyfert~2 galaxies harbour hidden BLRs, constraining both the ubiquity of obscuration and the geometry of the scattering medium (see e.g. \citealt{barth1999a,barth1999b,moran2000,tran2003,ramosalmeida2016} and references therein).

Within the standard unification model of AGN, polarimetry is a consolidated and powerful tool to probe the inner parsec in type-1 systems: the proximity of the polarising medium to the AGN means that polarised light can carry the imprint of spatially resolved information on physical scales that are impossible to resolve
in total light. Optical spectropolarimetry of Seyfert~1 galaxies reveals diverse polarisation behaviours across broad lines and continua that can be understood in terms of competing equatorial and polar scattering components, with the observed polarisation degree and position angle strongly dependent on inclination \citep{smith+2002,smith+2004,goosmann+gaskell2007}. Moreover, the wavelength-dependent polarisation angle and degree across the broad Balmer lines encode BLR kinematics and geometry, enabling complementary constraints on BLR structure \citep{smith+2002,smith+2004} and even single-epoch black-hole mass measurements \citep{afanasiev+2015}. Detailed equatorial-scattering models further show that large-amplitude blue-to-red swing in polarisation angle and wing peaks in polarisation fraction $p$ arise naturally when a rotating line-emitting disc is surrounded by a compact coplanar scattering region \citep{smith+2005}.

At the same time, a polarised optical continuum does not necessarily imply a distinct external scattering region. Since the optical/UV continuum of AGN is widely attributed to accretion-disc emission, intrinsic continuum polarisation has long been expected from radiative transfer within the disc atmosphere, driven primarily by electron scattering. Plane-parallel electron-scattering calculations predict
substantial polarisation at high inclination, although subsequent work showed that bound-free opacity, relativistic effects, non-planar disc geometries, and magnetic Faraday rotation can strongly alter both the degree and position angle of the emergent polarisation \citep{laor+1990,coleman+shields1990,kartje+konigl1991,agol+blaes1996}. A polarised AGN continuum therefore does not by itself require that the dominant polarisation arises in a separate external scattering medium, by which we mean material outside the continuum-emitting region, i.e. the accretion disc.

In this article we present optical linear spectropolarimetry of the local LRD \target. After presenting the observations and data reduction (Section~\ref{s.data}), we show the results of our analysis (Section~\ref{s.r}) and discuss the implications (Section~\ref{s.d}). We conclude with a summary of our findings in Section~\ref{s.conc}. Throughout this article, we assume the \textLambda CDM cosmology from the \citet[][their table~2]{planck+2020}. All wavelengths are air wavelengths, and line equivalent widths (EW) are in rest frame, with negative values corresponding to line emission. We adopt the standard IAU convention that the polarisation angle $\vartheta$ is measured from North to the East; we define the polarisation reference such that $(q,u) = (1,0)$ and $(q,u) = (0,1)$ denote polarisation along $\vartheta=0$\textdegree\ and $\vartheta=45$\textdegree, respectively.

\section{Observations and data reduction}\label{s.data}

We present optical linear spectropolarimetry of \target with VLT/FORS2 in PMOS mode, obtained by the dedicated ESO programme 115.29FY.001. The observations consisted of eight successful observing blocks (OBs), taken between UT~2025-12-23 and 2026-02-22. We rejected three additional OBs due to low signal-to-noise ratio (SNR) in one or more configurations. The on-chip seeing spanned a full-width at half maximum (FWHM) of 0.7--0.89 arcsec, as measured by fitting a Gaussian to the 2-d spectrum of the broad \Halpha wings, which we assumed to be spatially unresolved (unlike narrow \Halpha; Section~\ref{s.r.ss.spatial}). The PMOS setup used a 1-arcsec-wide slit and the GRIS~1200R grating, covering $5750<\lambda<7310\,\AA$ at a resolution $R\simeq2140$ at \Halpha. This nominal resolution is consistent with the on-chip determination, where we find a $FWHM=3.083~\AA$ at \Halpha ($R=2128$), obtained from fitting sky lines and interpolating linearly in wavelength to reach \Halpha. We used the GG435 order-sorting filter and $2\times2$ on-chip binning to improve the SNR in the continuum and broad-line wings. Targeting linear polarisation, each OB was split into four retarder-plate position angles $\alpha = 0$\textdegree, 22.5\textdegree, 45\textdegree, and 67.5\textdegree.

We reduced the data using the ESO \textsc{Reflex} standard pipeline for FORS2
spectropolarimetry, which performs bias subtraction, flat-field correction,
wavelength calibration, sky subtraction, extraction of the ordinary and
extraordinary beams, and derivation of Stokes $Q(\lambda)$ and $U(\lambda)$.
The eight retained OB spectra were resampled onto a common wavelength grid and
combined by computing the cross-OB median of $Q(\lambda)$, $U(\lambda)$, and
$I(\lambda)$. We experimented with arithmetic means, clipped means, and
inverse-variance weighting, but the median stack provided the cleanest
suppression of residual artefacts while preserving the same qualitative line
effects. Unless stated otherwise, all measurements refer to this
median stack.
We define the normalised Stokes parameters $q\equiv Q/I$ and $u\equiv U/I$,
the fractional linear polarisation $p\equiv\sqrt{q^2+u^2}$, and the
quadrant-preserving polarisation angle $\vartheta \equiv 0.5 \arctan(u,\,q)$.
The linearly polarised (hereafter simply `polarised') flux is then the product
$p\times I = \sqrt{U^2 + Q^2}$.

\begin{figure}
\centering
\includegraphics[width=\columnwidth]{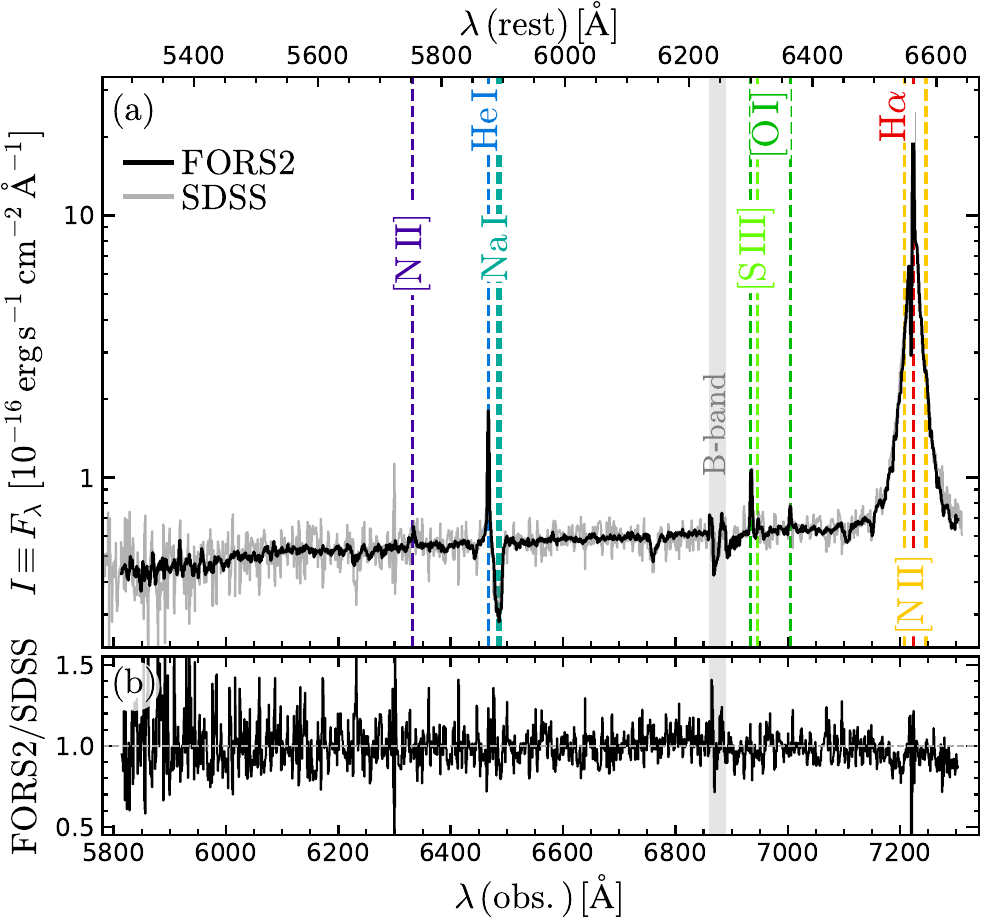}
\caption{Flux-calibrated FORS2 spectrum of \target (black) compared with the
archival SDSS spectrum (grey). The lower panel shows any residual
flux-calibration mismatch. The grey shaded band marks the
telluric B absorption band. The two spectra agree well across most of the
wavelength range; a localised flux deficit in FORS2 relative to SDSS is visible
just blueward of \Halpha.}\label{f.data}
\end{figure}

The stacked flux spectrum $I$ is shown in Fig.~\ref{f.data}; we used the
SDSS Data Release~7 spectrum of the source \citep{abazajian+2009} for absolute
flux calibration. Our wavelength solution yields a redshift of $z=0.100645$;
after converting our FORS2 air wavelengths to vacuum to match SDSS, the
residual offset from the SDSS redshift is $-34$~\kms. This is not due to
our frame of reference, since both works use barycentric redshift, so we
attribute this offset to systematics in our wavelength calibration. To avoid
additional resampling, we adopt the FORS2 wavelength scale for the rest of
this article.
After correcting for this velocity shift, the FORS2 and SDSS spectra agree
well across the full FORS2 range (Fig.~\ref{f.data}), with the exception of
the atmospheric B-band absorption near $6870~\AA$.
We noticed a localised flux depression in the FORS2 data just blueward of
\Halpha ($\sim7200~\AA$ observed), without a counterpart in SDSS; given the
variability in both emission and absorption-line equivalent width tentatively
reported in LRDs at $z=5\text{--}7$ \citep{deugenio+2026a,ji+2025a,furtak+2025a},
but in light of the limited wavelength-calibration accuracy in this detector
region, we simply note the feature but do not interpret it further.

To validate our results, we inspected per-OB $q\text{--}u$ profiles and
polarised-flux spectra; we repeated our analysis by using a fixed box-car
extraction (16 spatial pixels) instead of the default optimal extraction,
and we used leave-one-out and leave-two-out stacking. For the latter test,
removing any one OB does not affect our results; for example the changes in
the median continuum or line polarisation are $\lesssim0.2$~per cent (the
largest change being $0.23$~per cent).

The sightline towards \target is relatively free of Galactic dust, with
$A_{V,\rm MW}=0.14$~mag \citep{schlafly+finkbeiner2011,ji+2026a}. Using the relation of
\citet{serkowski+1975}, this value implies an upper limit to the foreground
(Milky-Way) polarisation of $p_\mathrm{MW} < 3 \times (A_{V, \rm MW}/ \mathrm{mag})$
per~cent, or 0.4~per cent. We additionally assessed the residual instrumental
polarisation using nightly observations of unpolarised standard stars, finding
$p_{\rm inst}\lesssim0.2$~per cent. While non-negligible, these contributions
are unlikely to account for the polarisation measured in our target
(Section~\ref{s.d.ss.intrinsic}). In the remainder of the article, we do not
subtract these contributions, but we discuss where they may be relevant and
affect the interpretation of the data.

\section{Results}\label{s.r}

\subsection{A polarised continuum}\label{s.r.ss.polcont}

\begin{figure*}
{\phantomsubcaption\label{f.specpol.a}
\phantomsubcaption\label{f.specpol.b}
\phantomsubcaption\label{f.specpol.c}
\phantomsubcaption\label{f.specpol.d}
\phantomsubcaption\label{f.specpol.e}
\phantomsubcaption\label{f.specpol.f}
\phantomsubcaption\label{f.specpol.g}
\phantomsubcaption\label{f.specpol.h}
\phantomsubcaption\label{f.specpol.i}
\phantomsubcaption\label{f.specpol.j}
\phantomsubcaption\label{f.specpol.k}
\phantomsubcaption\label{f.specpol.l}}
\centering
\includegraphics[width=\textwidth]{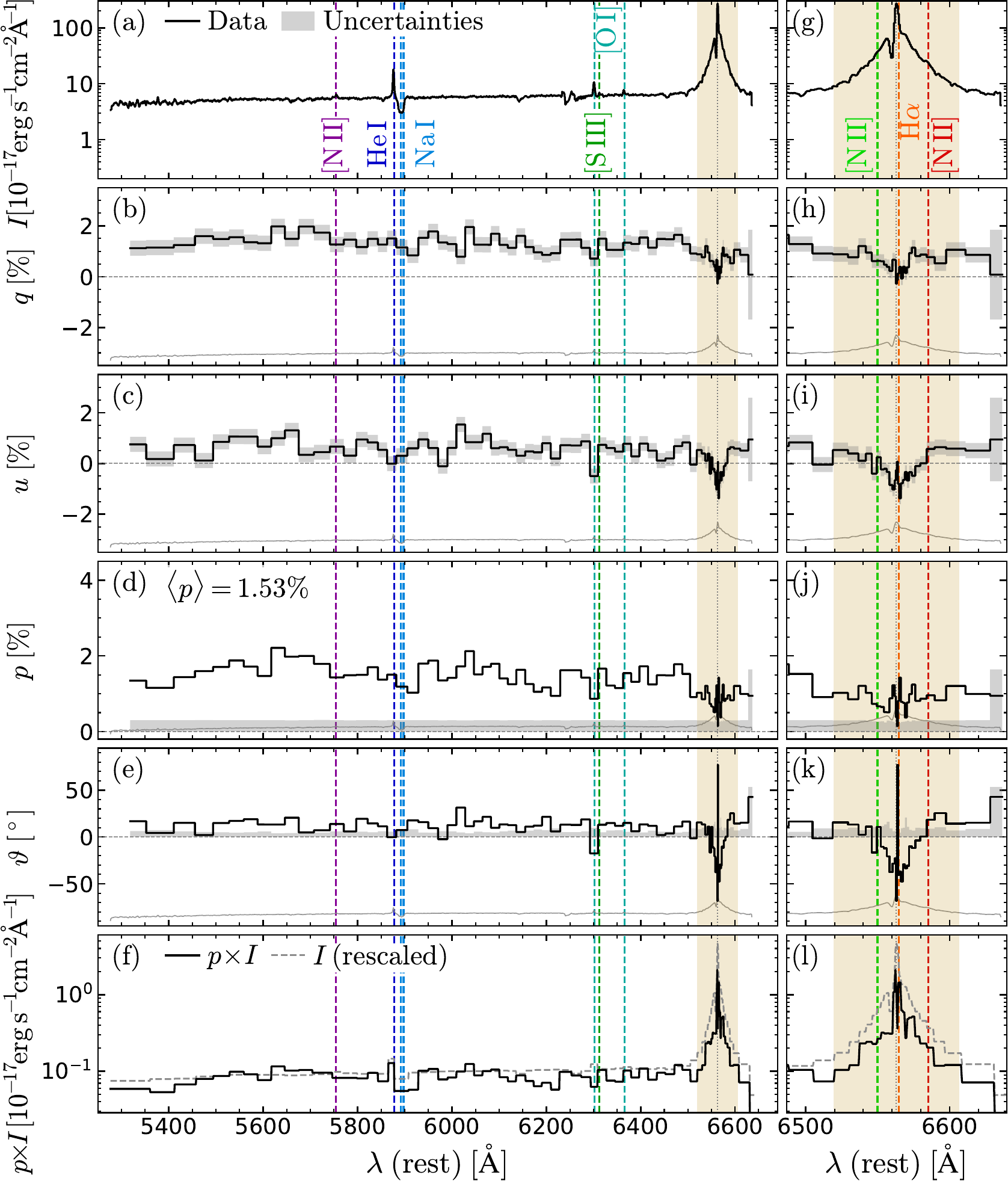}
\caption{Optical linear spectropolarimetry of \target (binned adaptively to
a fixed $\sigma(p)\simeq0.3$\,per cent, using the cross-OB median stack). The
shaded vertical region encloses $\pm2000$~\kms from systemic \Halpha. The
continuum displays nearly grey polarisation, with $\langle p \rangle = 1.53
\pm0.04$(rand.)$\pm0.20$(syst.) per cent and a consistent $\vartheta=
+12$\textdegree over the available wavelength range (panel~\subref{f.specpol.d}),
while broad \Halpha is even less polarised (panel~\subref{f.specpol.d} and
zoom-in panel~\subref{f.specpol.j}). We detect a swing in the polarisation
angle between the continuum (near $\vartheta \simeq +12$\textdegree;
panel~\subref{f.specpol.e}) and broad \Halpha, which reaches
$\vartheta\simeq-36$\textdegree, dipping locally to $\simeq-40$\textdegree\ (panel~\subref{f.specpol.k}).
The continuum polarisation is significantly detected and larger than any
systematic effects such as foreground polarisation and residual instrumental effects
(Section~\ref{s.data}). The polarised flux $p\times I$
(panel~\subref{f.specpol.f}) follows the shape of total flux $I$ (grey dashed
line), whereas broad \Halpha is clearly suppressed in polarised flux
(panel~\subref{f.specpol.l}).
}\label{f.specpol}
\end{figure*}

Fig.~\ref{f.specpol} summarises the optical spectropolarimetry data; the \Halpha
region is highlighted in sand colour (we show a zoom in the right panel). Stokes $I$ is
shown in the native binning, while all other panels are binned to overcome Rice
bias in $p$ \citep{wardle+kronberg1974} and periodic instabilities in the
polarisation position angle $\vartheta$.
To preserve spectral resolution where the SNR allows it, we use an adaptive
binning scheme, tailored to reach a target of 0.3~per cent noise in $p$.

The spectrum is clearly polarised (panel~\subref{f.specpol.d}), as also evident
from the non-zero values of both $q$ and $u$, ruling out any residual Rice bias
after our adaptive binning (panels~\subref{f.specpol.b} and~\subref{f.specpol.c}).
The polarisation value
is fairly low, averaging $\langle p \rangle = 1.53\pm0.04$~per cent over the
entire spectral window, after masking out \Halpha, the weaker spectral
features (\HeIL, Na~D, \OIall), and the telluric B band. This value is comparable
to what found in many local Seyfert 1 galaxies \citep[e.g.,][]{smith+2002,
smith+2004,afanasiev+2019}.

\subsection{Grey polarisation curve}\label{s.r.ss.polcol}

\begin{figure}
{\phantomsubcaption\label{f.polcol.a}
\phantomsubcaption\label{f.polcol.b}
\phantomsubcaption\label{f.polcol.c}
\phantomsubcaption\label{f.polcol.d}}
\centering
\includegraphics[width=\columnwidth]{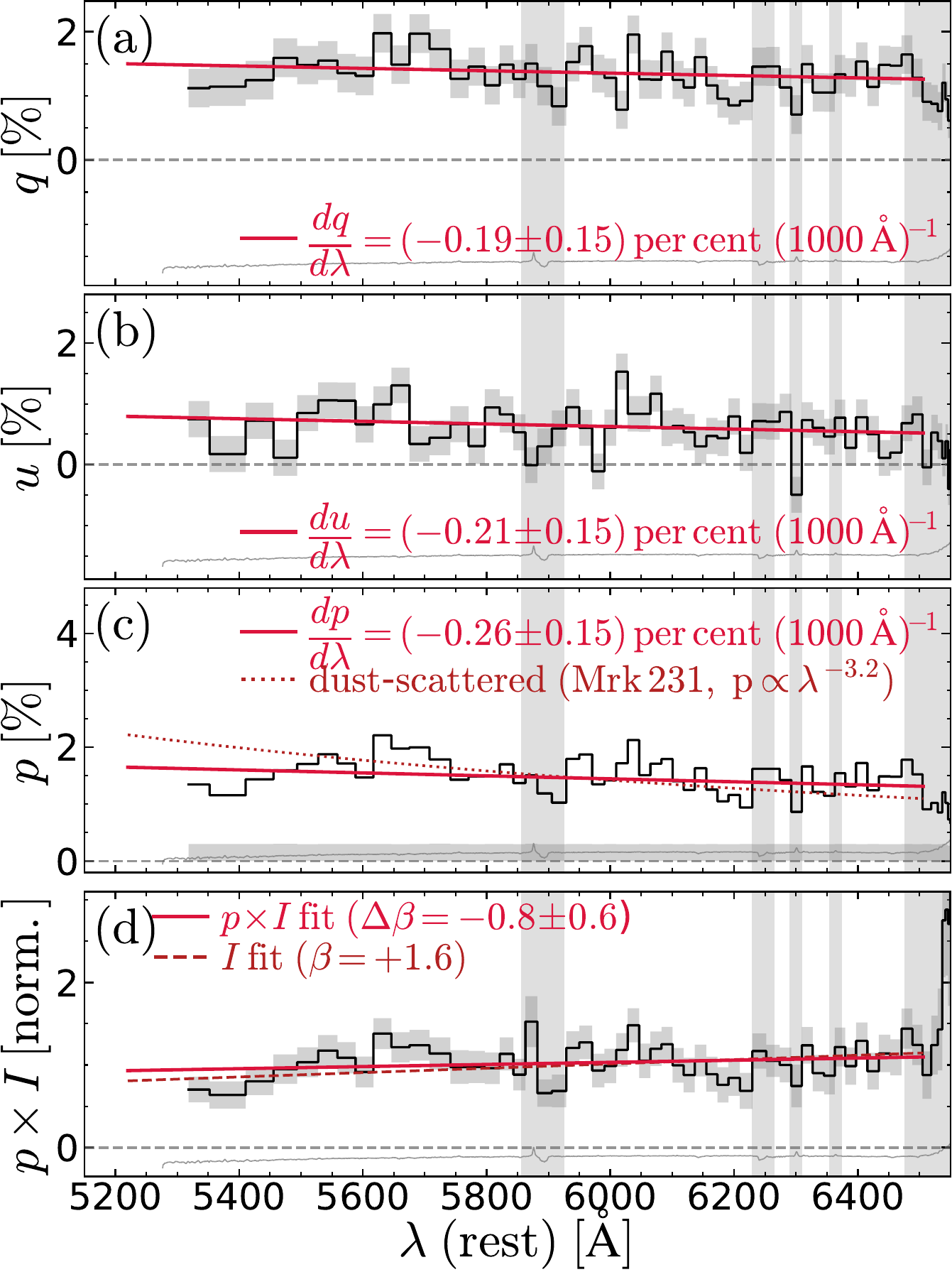}
\caption{Continuum polarisation slope fit over the line-free continuum (the grey
vertical bands are masked). In each panel, the black line and grey shaded band
are the data and uncertainties, while the best-fit model is in red.
Panels~\subref{f.polcol.a} and~\subref{f.polcol.b} show direct fits
to $q$ and $u$. Panel~\subref{f.polcol.c} shows the slope derived from combining
the models for $u$ and $q$, compared with the $p(\lambda)$ expected for
Markarian-231-type dust scattering \citetext{$p\propto\lambda^{-3.2}$,
dotted line; \citealp{smith+1995}}, which is clearly inconsistent
with the data.
Panel~\subref{f.polcol.d} is a fit to the polarised flux (normalised at
5900~\AA), reporting the slope difference $\Delta\beta$ with respect to the
flux $I$, adopting the convention $F_\lambda\propto\lambda^\beta$; our data and
the resulting best-fit model are consistent with grey
scattering (dashed grey, the linear fit to $I$; electron scattering predicts
$\Delta\beta=0$).}\label{f.polcol}
\end{figure}

The spectral dependence of line polarisation carries information on the conditions
of the source, and on the nature of the polarising medium. For example, electron
scattering is wavelength independent (grey polarisation), while dust scattering
is strongly wavelength dependent \citep[e.g.,][]{miller+1991,goosmann+gaskell2007}.
To test the wavelength dependence of $p$, we fitted local linear slopes to the
continuum Stokes spectra over the full line-free continuum (Fig.~\ref{f.polcol}),
masking broad \Halpha ($\pm4000~\kms$), \HeIL\ and Na~D, the \OIall doublet, and
the telluric B band (rest-frame 6228--6264~\AA). Once again, we must control for
Rice bias, since the noise level increases at short wavelengths.
We therefore derive $dp/d\lambda$ from the individual Stokes slopes. From the
chain rule of $p \equiv \sqrt{q^2+u^2}$ we obtain
\begin{equation}
\frac{dp}{d\lambda} = \frac{1}{p} \left( q \, \dfrac{dq}{d\lambda} + u\,\dfrac{du}{d\lambda} \right) \;.
\label{eq:dpbias}
\end{equation}
We verified that this Stokes-space estimate is stable across all binning factors 1, 9, 16 and $25\times$ (varying by $\lesssim0.1$ per cent per 1000~\AA). In contrast, the direct slope fit from unbinned $p$ is biased steep by a factor of 2.4, as expected from Rice bias disproportionately affecting blue wavelengths. We find only a weak decline in the continuum polarisation, with $dp/d\lambda = -0.26 \pm 0.15$ per cent per 1000~\AA\ rest-frame (Fig.~\ref{f.polcol}); all wavelength slopes quoted below are likewise given per rest-frame units. This result is driven by $dq/d\lambda = -0.19\pm0.15$ per cent per 1000~\AA (panel~\subref{f.polcol.a}) and $du/d\lambda = -0.21 \pm 0.15$ per cent per 1000~\AA (panel~\subref{f.polcol.b}), while considering that $q>0$ and $u>0$ over the whole wavelength range. A direct fit to the adaptively binned $p$ spectrum gives a consistent result, $dp/d\lambda = -0.26\pm0.15$ per cent per 1000~\AA. The signal of a spectral gradient in the polarisation of the continuum is therefore marginal in the median-stack analysis, hovering at $\simeq 1.7~\sigma$ significance (the mean-stacked analysis returns a steeper, 3.5-\textsigma\ slope, Appendix~\ref{a.altmean}).

Panel~\subref{f.polcol.d} shows the spectral slope of the polarised flux
$p\times I$. For $I$, assuming $I(\lambda) \propto \lambda^\beta$, we measure
$\beta = +1.6$ \citep[in agreement with ][]{ji+2026a}. For $p \times I$, we
measure a slope steepening $\Delta\beta \equiv
\beta_{p\times I} - \beta_I = -0.80\pm0.57$, i.e., the polarised continuum
is statistically consistent with having the same spectral slope as the total flux,
implying no significant evidence of wavelength dependence from the optical data alone. Clearly,
a larger wavelength range is required to obtain a decisive proof of a
polarisation gradient.

The mild redward slope in $p(\lambda)$ established above (Fig.~\ref{f.polcol})
implies that the continuum polarisation declines at only $\simeq0.3$\,per cent
per 1000~\AA, so the optical continuum is consistent with being approximately
grey, at least over our limited $\sim$1300-\AA baseline in wavelength.
We note that, because the empirical dichroic curve of Galactic dust is broad
\citetext{shape parameter $K\simeq1.15$; \citealp{serkowski+1975}} and
our baseline is short, a dichroic $p(\lambda)$ peaking near our observed range
would itself appear nearly grey, varying by $\lesssim0.1$~per cent across our
baseline; our polarisation curve therefore does not discriminate against
dichroic transmission.

\begin{table}
\centering
\caption{Polarisation and ancillary measurements of \target. All per-wavelength
units are rest-frame. Polarisation fractions are quoted as
value\,$\pm$\,statistical\,$\pm$\,systematic, where the systematic is the
instrumental polarisation $p_{\rm inst}\lesssim0.2$ per cent (Section~\ref{s.data}).
The continuum-subtracted line quantities ($p_{\Halpha}$, $\vartheta_{\Halpha}$)
are less sensitive to a constant Stokes-flux offset, but the exact line-only
polarisation angle can still be shifted by a residual instrumental polarisation term
because $p_{\Halpha}$ is small (Section~\ref{s.r.ss.hapol}).
}\label{t.measurements}
\begin{tabular}{ll}
\hline
Quantity & Value \\
\hline
Continuum      &                               \\
$p_{\rm cont}$ & $1.53\pm0.04\pm0.20$ per cent \\
$\vartheta_{\rm cont}$ & $+12\pm1$\textdegree  \\
$dp/d\lambda$ & $-0.26\pm0.15$ per cent per $(1000~\AA)^{-1}$ \\
$\Delta\beta$ & $-0.8\pm0.6$ \\
$p_{\rm cont}/A_{V,\rm MW}$ & $10.9$ per cent~mag$^{-1}$ \\
\hline
Broad \Halpha (cont. subtracted) & \\
$p_{\Halpha}$ & $0.58\text{--}0.84$ per cent \\
$\vartheta_{\Halpha}$ & $-34\pm3$\textdegree \\
$\Delta\vartheta=\vartheta_{\Halpha}\!-\!\vartheta_{\rm cont}$ & $-48\pm4$\textdegree \\
\hline
\end{tabular}
\end{table}

\subsection{Line polarisation}\label{s.r.ss.polline}

Across the brightest emission lines we detect line effects, i.e., localised
changes in the polarisation degree $p$ and/or position angle $\vartheta$,
confined to a line relative to the adjacent continuum
\citep[e.g.,][]{antonucci+miller1985,smith+2002}. Such effects are present at
\HeIL, \OIL, and both the broad and the narrow components of \Halpha: in both
$q$ and especially $u$ these lines deviate from the continuum
(Fig.~\ref{f.specpol.b} and~\ref{f.specpol.c}). This is also visible as a
depolarisation signal (localised drop in $p(\lambda)$) at the location of some
of these lines, most notably \OIL and \Halpha. It is
worth distinguishing two cases. Pure depolarisation by an intrinsically
unpolarised line lowers $p$ but preserves $\vartheta$; a change in $\vartheta$
instead requires the line to carry its own polarisation at a different angle, and
cannot arise from dilution alone. We see both: \HeIL, despite its higher
equivalent width, keeps the continuum angle (consistent with simple dilution,
and showing no significant $u$ jump), whereas \OIL shows a highly significant
jump in $u$ and a swing to negative $\vartheta$, echoing broad \Halpha
(Fig.~\ref{f.specpol.e}).

Compared to the continuum, broad \Halpha emission is significantly less polarised, with a central minimum near line peak reaching $\simeq$0.5~per cent, approaching
the ceiling of our systematic error ($<0.2$~per cent). This drop in $p$ is driven
by both polarisation components: $q$ tends to 0 for wavelengths approaching line centre, while $u$ crosses 0 to reach opposite values with respect to the continuum average.

In principle, line depolarisation could be due to dilution of the polarised continuum by a completely depolarised emission line.
Specifically, pure depolarisation would naturally explain the observed drop in $q$; it cannot, however, explain the change of sign in $u$, thus indicating that the \Halpha line must have its own non-zero polarisation with a different angle with respect to the continuum. In mathematical terms, we can illustrate this with a simple model of flux-weighted mixing in Stokes space. If the continuum and line have Stokes parameters $(\qc,\uc)$ and $(\ql,\ul)$, then
\begin{equation}
q_\mathrm{obs} = \frac{\qc F_\mathrm{c} + \ql F_{\Halpha}}{F_\mathrm{c} + F_{\Halpha}}, \quad
u_\mathrm{obs} = \frac{\uc F_\mathrm{c} + \ul F_{\Halpha}}{F_\mathrm{c} + F_{\Halpha}}, \quad
\label{eq.mixing}
\end{equation}
Therefore, the observed drop in polarisation across \Halpha cannot be fully explained by flux dilution due to an unpolarised line, but must be sustained by low-level polarisation of the line itself.

We also rule out that the drop in $p$ across \Halpha can be explained by a smooth continuum trend, since $q$ clearly rises again after the line trough (Fig.~\ref{f.specpol.i}), and $u$ even rises to the same magnitude as blueward of \Halpha.

Confirming the hypothesis of a non-zero position-angle offset ($\Delta\vartheta \neq 0$ above), we detect a change in polarisation angle between the constant value seen in the continuum, $\vartheta_{\rm cont}\simeq +12$\textdegree, and \Halpha. The wavelength trends in both $u$ and $q$ show that the change is genuine, although the actual angle is strongly diluted by the continuum Stokes vector (see Section~\ref{s.r.ss.hapol} for the continuum-subtracted angle).

\begin{figure}
\centering
\includegraphics[width=\columnwidth]{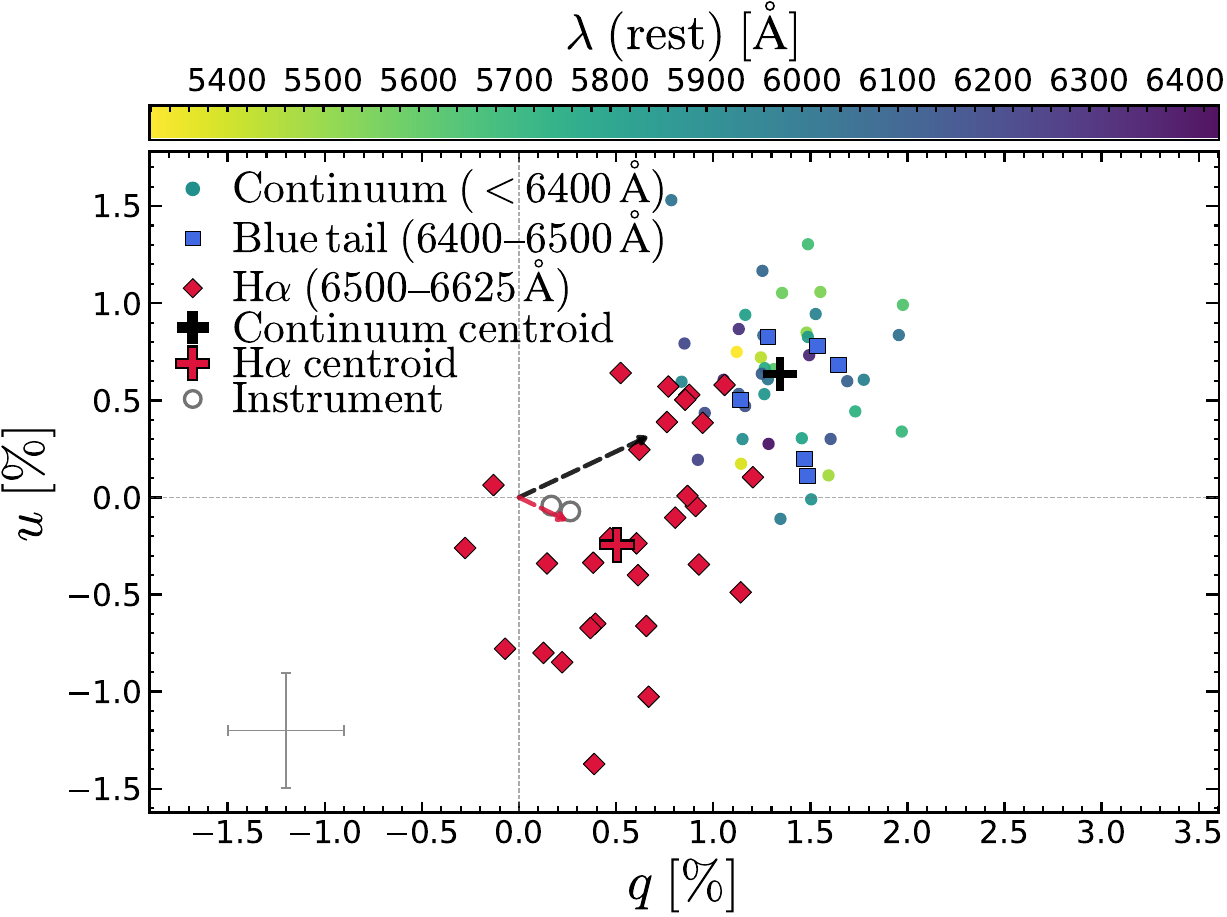}
\caption{$q\text{--}u$ plane diagram for the adaptively binned stacked spectrum. Small dots, coloured by rest-frame wavelength, show the continuum ($\lambda<6400$~\AA; bins falling in the telluric B band and on the \OIall doublet are excluded). Larger markers highlight the blue-tail (6400\text{--} 6500~\AA) and \Halpha (6500\text{--}6625~\AA) windows. The black and red crosses mark the continuum and \Halpha centroids (inverse-variance-weighted means). The grey empty circles show the instrument polarisation, which we do not remove. \Halpha shows a modest centroid offset and increased scatter relative to the continuum locus. This supports the presence of a weaker but non-zero Stokes component in the line.}\label{f.quplane}
\end{figure}

Non-null line polarisation is also evident in the $q\text{--}u$ plane (Fig.~\ref{f.quplane}). The blue-tail points are offset towards the origin, consistent with both lower polarisation and dilution, but \Halpha shows a clear centroid offset from the continuum direction, confirming the offset in polarisation angle and ruling out pure dilution. There is also a clearly visible increase in scatter relative to the continuum points, despite the measurement uncertainties being the same, due to the adaptive binning. Note that the points do not form a rotation-driven loop, but are instead consistent with a weak additional Stokes component in the line, while the scatter may be indicating a more structured or multi-component \Halpha-forming region \citep[e.g.,][]{ji+2026b}. Taken together, the total line spectropolarimetry shows that both the continuum and broad \Halpha are polarised; while less polarised than the continuum, the polarisation of the broad line cannot be explained solely by continuum dilution due to an unpolarised line, so a weak intrinsic broad-line Stokes component is required.

\subsection{Continuum-subtracted \texorpdfstring{\Halpha}{Ha}polarisation}\label{s.r.ss.hapol}

\begin{figure}
\centering
{\phantomsubcaption\label{f.contsubtheta.a}
\phantomsubcaption\label{f.contsubtheta.b}
\phantomsubcaption\label{f.contsubtheta.c}
\phantomsubcaption\label{f.contsubtheta.d}}
\includegraphics[width=0.92\columnwidth]{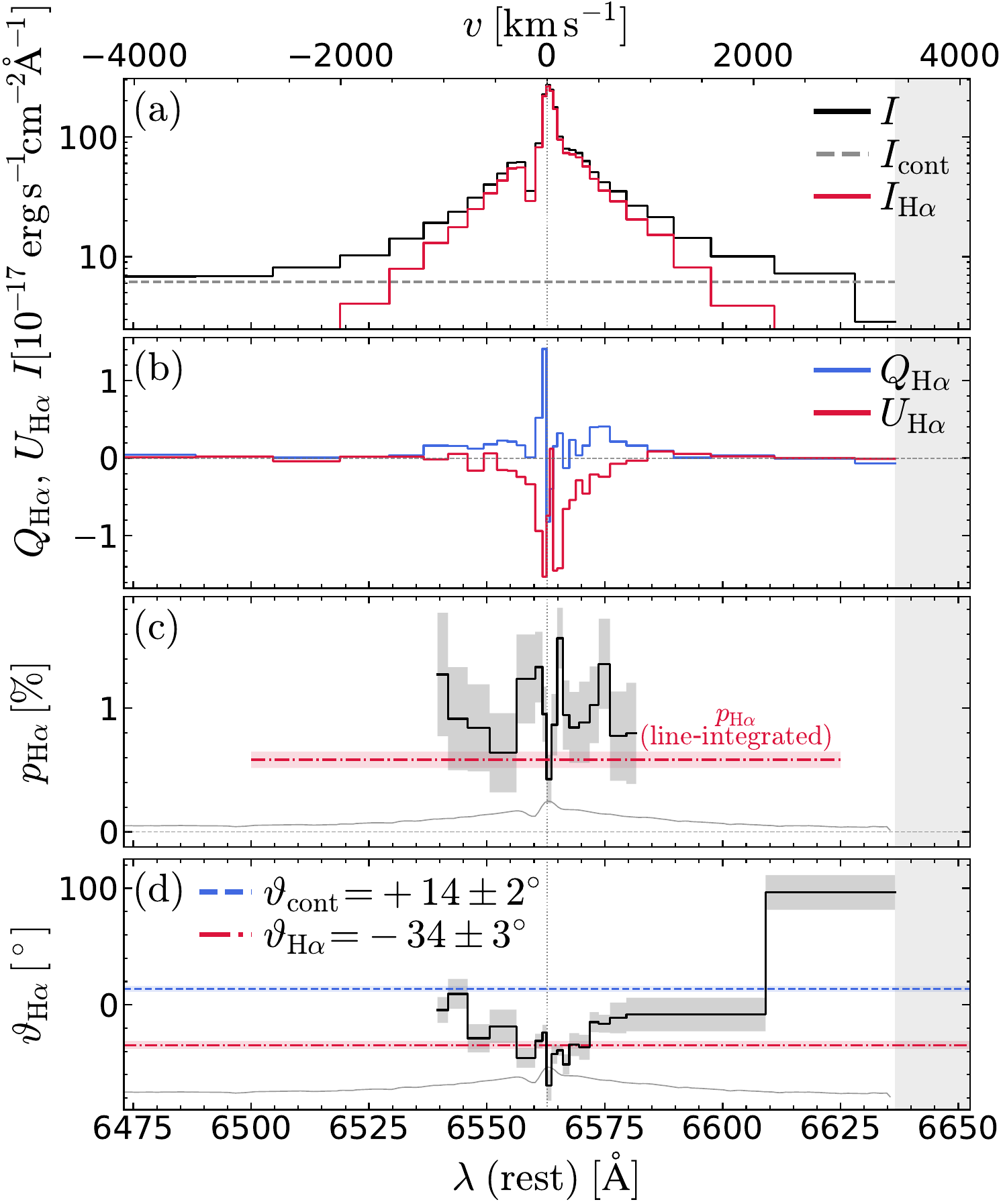}
\caption{Continuum-subtracted Stokes-angle for broad \Halpha in the adaptive-binned median stack. We subtract the constant continuum Stokes vector (the mean of the binned Stokes-flux spectra $I$, $Q=qI$, and $U=uI$ over 6200\text{--}6400~\AA), before recomputing $p_{\Halpha}$ and $\vartheta_{\Halpha}$. The top panels show the continuum and residual line-only Stokes fluxes. Panel~\subref{f.contsubtheta.c} shows the continuum-subtracted $p_{\Halpha}(\lambda)$; the dash-dotted line and band mark the `coherent' (line-integrated) polarisation, $p_{\Halpha}=0.58\pm0.24$ per cent; note the per-bin values are positive-definite moduli, insensitive to the angle rotation, so they lie systematically above the line-integrated value (Section~\ref{s.r.ss.hapol}). Panel~\subref{f.contsubtheta.d} shows the continuum-subtracted position angle. A robust line angle offset is present, without a blue-to-red swing in $\vartheta_{\Halpha}$. The grey shading are beyond the spectral coverage.}\label{f.contsubtheta}
\end{figure}

The \Halpha line has a lower polarisation fraction than the continuum ($p_{\Halpha}< p_{\rm c}$), but the absolute polarised flux is still enhanced at \Halpha: Fig.~\ref{f.specpol.f} shows that near the location of the \Halpha peak, $p\times I$ is $\sim 4$ times higher than the continuum polarised flux. Of course, this is due to the line's large total flux, which amplifies even a
small polarisation fraction. To quantify the intrinsic polarised line signal more directly, we subtract a constant continuum Stokes component, measured as the mean of the adaptively binned Stokes-flux spectra $I$, $Q=qI$, and $U=uI$ over 6200--6400~\AA (with the same masks as Section~\ref{s.r.ss.polcont}). All line-only quantities below derive from the resulting binned, continuum-subtracted Stokes fluxes. The residual line-only Stokes profiles (Fig.~\ref{f.contsubtheta.b}) show a coherent signal: 
$Q_{\Halpha}$ is systematically positive across the entire profile, whereas $U_{\Halpha}$ is mostly negative, but with a narrow positive spike at the line core. We then condense these profiles into two moments over the \Halpha window (6500\text{--}6625~\AA; Fig.~\ref{f.quplane}), which contains almost the entirety of the line flux: the wavelength-integrated polarisation, i.e., the coherent modulus of the summed Stokes fluxes, and the wavelength-average polarisation, i.e., the median of the per-bin $p$ (Table~\ref{t.measurements}). The first yields $p_{\Halpha}=0.58\pm0.24$ per cent, the second $p_{\Halpha}=0.84$~per cent (larger because the per-bin moduli are defined positive and blind to the angle oscillations leading to partial vector cancellation in the coherent sum). Hence we use $p_{\Halpha}=0.58\text{--}0.84$~per cent as the fiducial range. From bootstrapping, we obtain a 95\textsuperscript{th}-percentile upper bound of $p_{\Halpha}<0.96$ per cent, lower than the continuum average $p_{\rm c}=1.53$ per~cent.

Moving to the polarisation angle, the same continuum-subtracted Stokes fluxes (Fig.~\ref{f.contsubtheta}) yield a line-integrated angle $\vartheta_{\Halpha}=-34\pm3$\textdegree, offset by $\Delta\vartheta\equiv\vartheta_{\Halpha}-\vartheta_{\rm cont}\simeq-48$\textdegree from the continuum vector based on the same procedure ($\vartheta_{\rm cont}\simeq+14$\textdegree\ in the masked sideband; the full-range continuum angle is $+12\pm1$\textdegree, Table~\ref{t.measurements}). We detected no blue-to-red swing in the line.

As a foreground check, we repeated the same Stokes-space measurement after subtracting a constant Galactic Stokes vector with amplitude fixed either to a typical value, $p_{\rm MW}=0.2$~per cent, or to the maximum allowed by the sightline, $p_{\rm MW}=0.4$~per cent, while allowing its position angle to take any value. The offset changes for some foreground angles but is not removed: the minimum absolute offset is $\simeq39$\textdegree\ for $p_{\rm MW}=0.2$~per cent and $\simeq31$\textdegree\ for $p_{\rm MW}=0.4$~per cent. A residual instrumental polarisation at the level indicated by the zero-polarisation standards, $p_{\rm inst}\simeq0.2$--0.3~per cent, is formally equivalent to the foreground test and we can therefore conclude that the angular shift between line and continuum is robust also with respect to instrumental polarisation.

We warn that even in the adaptively binned data, both $\vartheta$ and $p$ display strong fluctuations near line centre. These could be due either to the fact that $p$ is small in this region, so $\vartheta$ is correspondingly less stable, and/or to the McLean effect \citep{mclean1979,ababakr+2017}, in which absorption near line core preferentially removes direct, unpolarised light, leaving the trough relatively enriched in scattered (and hence polarised) photons, thus raising the local polarisation.

\subsection{Constraints from the polarised-flux spectrum}\label{s.r.ss.polflux}

\begin{figure}
\centering
{\phantomsubcaption\label{f.pifit.a}
\phantomsubcaption\label{f.pifit.b}
\phantomsubcaption\label{f.pifit.c}}
\includegraphics[width=\columnwidth]{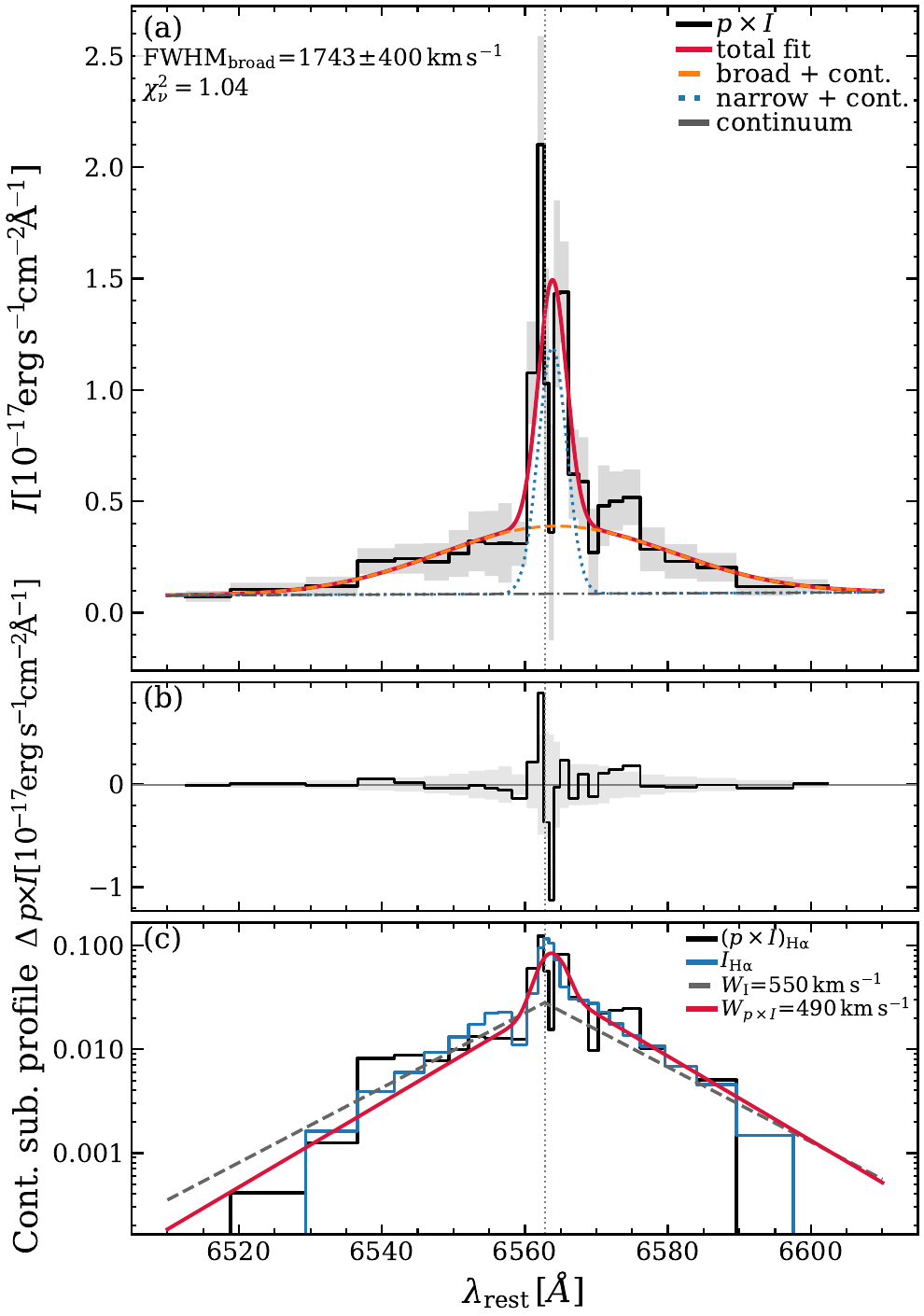}
\caption{Fit to the adaptive-binned polarised-flux spectrum $p\times I$ around \Halpha. Panel~\subref{f.pifit.a} shows the linear continuum plus narrow and broad Gaussian decomposition, yielding $FWHM_{\rm broad}(p\times I)=1740\pm400~\kms$ and $FWHM_{\rm narrow}(p\times I)=214\pm43~\kms$. Panel~\subref{f.pifit.b} shows the residuals. Panel~\subref{f.pifit.c} compares the continuum-subtracted $p\times I$ to the (rescaled) $I$: the two profiles overlap along the entire line. The solid red curve is the double-Gaussian $p\times I$ model, while the dashed grey curve overplots an exponential with $W=550~\kms$, as per our toy-model fit in total light (Section~\ref{s.r.ss.toyfit}). The two wing widths are statistically consistent with each other, but the uncertainties on $W_{p\times I}$ are large (see text).
}\label{f.pifit}
\end{figure}

To test whether the polarised-flux line profile is broader than the total-flux profile, as often seen in AGN lines subject to scattering by an external medium \citep[e.g.,][]{smith+2005,goodrich+miller1994}, we also fit the adaptively binned $p\times I$ spectrum with a linear continuum plus narrow and broad Gaussian \Halpha components (Fig.~\ref{f.pifit}). This fit yields $FWHM_{\rm broad}(p\times I)=1740\pm400~\kms$ and $FWHM_{\rm narrow}(p\times I)=214\pm43~\kms$. These are the widths of the two Gaussians fitted to the polarised flux, and should not be confused with the observed broad-line width in total light \citep[$FWHM_{\rm obs}=934\pm10~\kms$;][]{ji+2026a} or the intrinsic, deconvolved width in the electron-scattering model ($FWHM_{\rm int}=500\pm28~\kms$; Section~\ref{s.r.ss.toyfit}).

The broad polarised component is consistent with the broadest Gaussian of the two-component direct fit \citep[$FWHM\simeq2000~\kms$;][]{burke+2021,ji+2026a} and larger than the single-Gaussian total-intensity broad line. Given the large uncertainties, however, it is consistent within $1\text{--}3\sigma$ with essentially the entire range of total-intensity broad-line widths, from the $\simeq500~\kms$ intrinsic model value to the $\simeq2000~\kms$ broad Gaussian. Crucially, we do not find the markedly broader polarised line seen in some scattered-line AGN, where a narrow direct core coexists with much broader scattered wings \citep[e.g.,][]{miller+1991,goodrich+miller1994,laor2006}: as Fig.~\ref{f.pifit} (bottom) shows, the continuum-subtracted $p\times I$ and $I$ profiles, normalised to the same integrated area and binning, are mutually consistent.

To measure the exponential wing width of the polarised flux, we fit $p\times I$ with a linear continuum, a narrow Gaussian (the narrow polarised component above), and an exponential; the continuum is anchored to the line-free windows, where it is well determined. The fit yields $W_{p\times I}=490\pm130~\kms$, which is statistically consistent with the exponential width of the total-flux fit, $W_{I}=550\pm10~\kms$ (Section~\ref{s.r.ss.toyfit}).

The narrow polarised component ($214\pm43~\kms$) is, in turn, much broader than the genuine narrow \Halpha line \citep[$\sigma\simeq27\text{--}40~\kms$;][]{ji+2026a,lin+2026a}, which presumably traces the host galaxy (marginally resolved at a physical scale of $\sim0.5$~kpc, Section~\ref{s.r.ss.spatial}) and appears in polarised flux as a depolarisation dip, so it does not set the polarised width. The $214\pm43~\kms$ component is therefore incompatible with the host narrow line.

\subsection{Illustrative mixed direct-plus-scattered \texorpdfstring{\Halpha}{Halpha} fit}\label{s.r.ss.toyfit}

The addition of polarimetric data to the spectrum motivates a phenomenological test in which the observed profile is built from a dominant direct broad-line core plus a weaker, broader electron-scattered component. For Stokes-$I$ we use the same model as \citet[][as implemented in \citealt{deugenio+2025e,ji+2026a}]{rusakov+2026}, consisting of a linear continuum, a direct broad \Halpha component, and a broader scattered component with exponential wings. We also add a narrow-line complex and an absorbing component near line centre, which we model using the optical depth model \citetext{e.g., \citealt{rupke+2005}; see \citealp{deugenio+2026a,deugenio+2026b} for applications in the context of LRDs}. For $p$, we assume that the continuum is polarised with $p_\mathrm{cont}=1.51$~per cent (Section~\ref{s.r.ss.polcont}), the direct broad and narrow lines are completely unpolarised, while the broad-scattered line is polarised with free $p_{\Halpha}$. The model is then optimised against both the flux $I$ and $p$, in a narrow range within $-4000\text{--}3000~\kms$ from line centre.

\begin{figure}
{\phantomsubcaption\label{f.toyes.a}
\phantomsubcaption\label{f.toyes.b}
\phantomsubcaption\label{f.toyes.c}}
\centering
\includegraphics[width=\columnwidth]{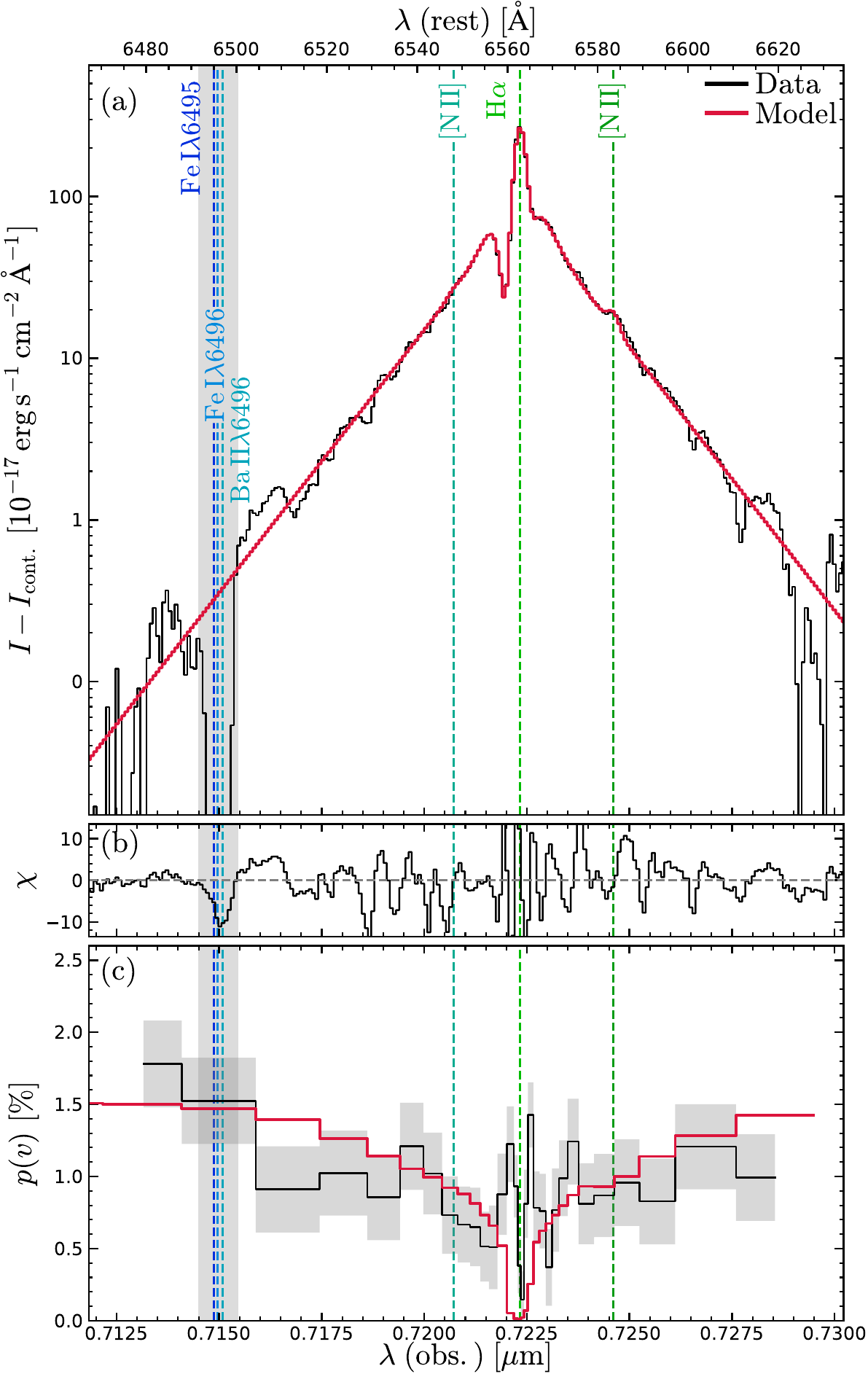}
\caption{
Illustrative toy model for a mixed direct-plus-scattered \Halpha profile. Panel~(a) compares the native-resolution median-stack total-flux spectrum (continuum-subtracted; black) with the best-fitting $I$-only model (red). Panel~(b) shows the corresponding residuals. Panel~(c) shows the resulting polarisation fraction $p$ across \Halpha (model, red; data, black), which reproduces the core depolarisation and the recovery of $p$ towards the continuum value in the wings. The model contains a direct broad \Halpha component, a broader electron-scattered component with approximately exponential wings, a narrow-line complex, and an absorbing component near line centre. Although the model correctly reproduces the overall trend of $I$ and $p$, significant residuals remain in both quantities, most notably close to the core of the line, indicative of an inherent complexity of the line forming region(s) for this object.}
\label{f.toyes}
\end{figure}

The resulting flux model is shown in Fig.~\ref{f.toyes.a}, and the best-fit parameters are listed in Table~\ref{t.toyes}, alongside the measurement uncertainties. We believe that the true uncertainties are dominated by systematic errors due to the limitations of the model, hence some of the formal values in the table may be underestimated. The extent of systematics can be gauged by inspecting the fit residuals (Fig.~\ref{f.toyes.b}), where high-significance undulations point to substantial data--model mismatch. This discrepancy may point to unaccounted for absorption and/or weak emission, and to a breakdown of some of the other assumptions, such as the fact that the narrow lines and absorption line are themselves unpolarised. However, delving into the absorption-line characterisation of LRDs is still challenging \citep[e.g.,][]{ji+2026a}, and is thus beyond the scope of the toy model and, first and foremost, of this paper too. With these limitations in mind, the best-fit parameters are in good agreement with the results of \citet{ji+2026a}, who based their analysis on twice-higher resolution spectroscopy from Gemini GMOS \citep{burke+2021}.
Specifically, we find $\tau_\mathrm{e} = 1.58\pm0.07$ against $1.79\pm0.05$,
our exponential kernel has width $W = 554\pm12~\kms$, while their value is
$569\pm4~\kms$, and the intrinsic BLR has $FWHM=500\pm28~\kms$, very close to
\citet{ji+2026a}'s $520\pm10~\kms$. Interestingly, we note that the
narrow-line width $\sigma_\mathrm{n} = 10\pm5~\kms$ is remarkably smaller
than in \citet{ji+2026a}, and more in line with the value of $\simeq 27~\kms$
from \citet[]{lin+2026a}, who used higher resolution still.
Among these model parameters, $\tau_\mathrm{e}$ provides independent constraints
on the amount of scattering underwent by \Halpha photons, which is crucial to
break the degeneracy between polarised fraction and geometry.

\begin{table}
\centering
\caption{Best-fit parameters for the toy electron-scattering
spectral model (see Fig.~\ref{f.toyes}).}
\label{t.toyes}
\begin{tabular}{lll}
\hline
Label & Value$\pm$Err & Units \\
\hline
$z_\mathrm{n}$ & $0.100649\pm0.000003$ & [---] \\
$\sigma_\mathrm{n}$ & $10\pm4$ & $[\mathrm{km\,s^{-1}}]$ \\
$F_\mathrm{n}(\mathrm{H\alpha})$ & $910\pm20$ & $[10^{-17} \, \mathrm{erg\,s^{-1}\,cm^{-2}}]$ \\
$F_\mathrm{n}(\mathrm{[N\,II]\lambda 6583})$ & $13\pm2$ & $[10^{-17} \, \mathrm{erg\,s^{-1}\,cm^{-2}}]$ \\
$FWHM_\mathrm{BLR}$ & $500\pm30$ & $[\mathrm{km\,s^{-1}}]$ \\
$F_\mathrm{b}(\mathrm{H\alpha})$ & $2670\pm40$ & $[10^{-17} \, \mathrm{erg\,s^{-1}\,cm^{-2}}]$ \\
$\tau_\mathrm{BLR}$ & $1.58\pm0.07$ & [---] \\
$W$ & $550\pm10$ & $[\mathrm{km\,s^{-1}}]$ \\
$v_\mathrm{abs}$ & $-50\pm2$ & $[\mathrm{km\,s^{-1}}]$ \\
$\sigma_\mathrm{abs}$ & $74\pm5$ & $[\mathrm{km\,s^{-1}}]$ \\
$\tau_{0,\mathrm{H\alpha}}$ & $3.2\pm0.6$ & [---] \\
$C_f$ & $1.00\pm0.07$ & [---] \\
$p_\mathrm{broad}$ & $0.8\pm0.5$ & $[\%]$ \\
\hline
\end{tabular}

\end{table}

Fig.~\ref{f.toyes.c} shows that this simple model reproduces the two key qualitative properties of the data at the same time: depolarisation in the line core and a recovery of the polarisation fraction toward the wings, where the model reaches the continuum value. Note that the polarisation reaches $\simeq 0$ at line centre due to the contribution of the unpolarised narrow component of \Halpha. We note that the fit is not statistically acceptable, and the parameters are strongly degenerate, so we do not regard it as a quantitative inference tool. In particular, the model strongly underpredicts $p$ close to the line core, pointing to complex behaviour near line resonance \citep[e.g.,][]{chang+2026}, clumpy structure \citep[e.g.,][]{ji+2026b}, or hard-to-model contribution from the host galaxy \citep[e.g.,][]{torralba+2026b,matthee+2026}. However, we also stress that the model--data mismatch is not driven by the polarised fraction alone: as discussed previously, significant shortcomings are also present in the flux model itself, and these are not driven by the simultaneous fit to $p$, since they persist with nearly identical shape and statistical significance when fitting Stokes $I$ alone (cf. Fig.~\ref{f.toyes.b} and Fig.~\ref{f.notoyes} from Appendix~\ref{a.notoyes}). In summary, this model provides a useful proof of concept for assessing the plausibility of some electron-scattered contribution to the broad-line wings. Finally, a double-Gaussian parametrization of the broad line would also reproduce the data \citep[e.g.,][]{ji+2026a}.

\subsection{Spatially resolved narrow \texorpdfstring{\Halpha}{Halpha}}\label{s.r.ss.spatial}

\begin{figure}
{\phantomsubcaption\label{f.spatialha.a}
\phantomsubcaption\label{f.spatialha.b}
\phantomsubcaption\label{f.spatialha.c}
\phantomsubcaption\label{f.spatialha.d}
\phantomsubcaption\label{f.spatialha.e}
\phantomsubcaption\label{f.spatialha.f}}
\centering
\includegraphics[width=\columnwidth]{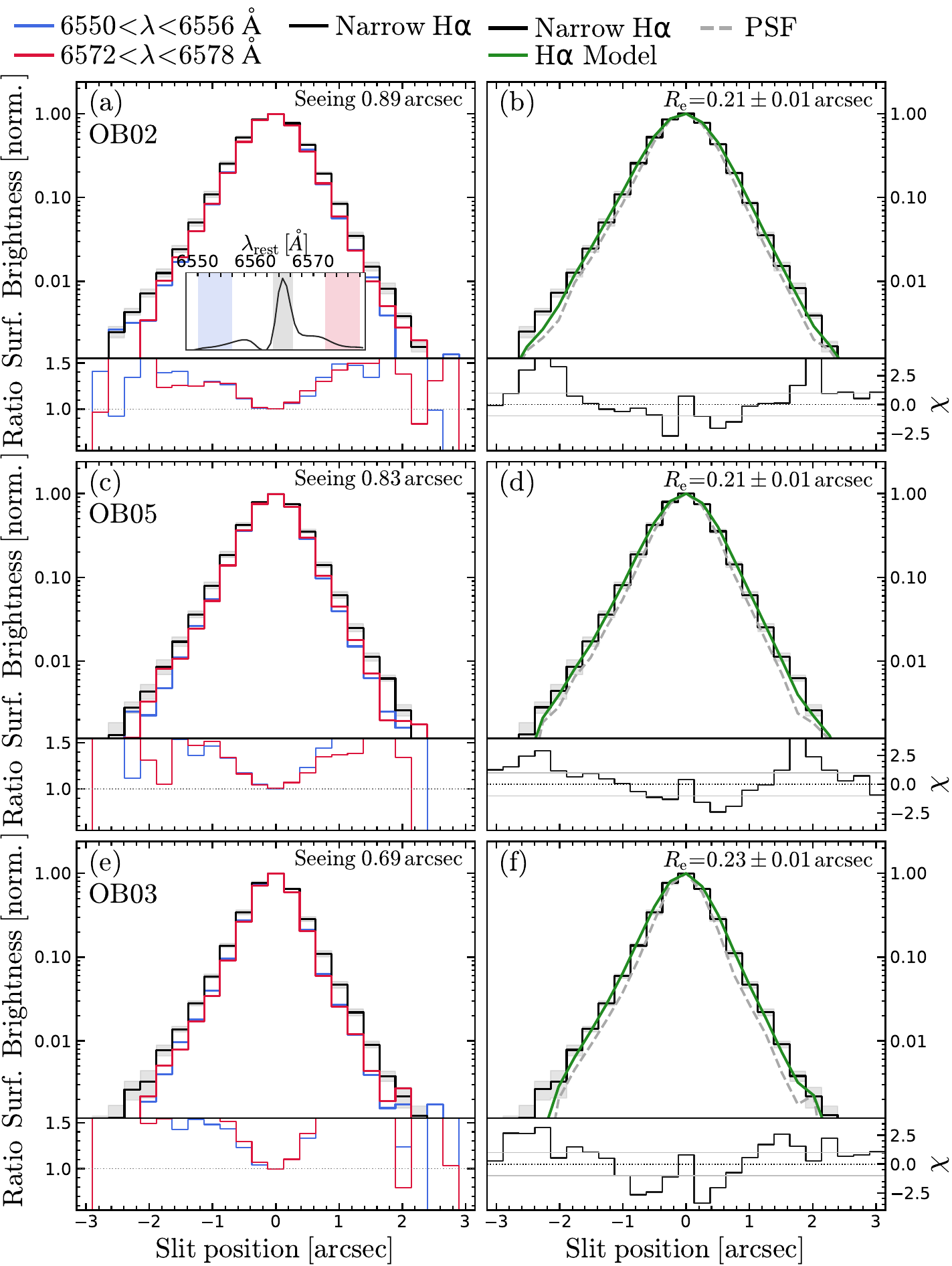}
\caption{Spatial profile comparison between narrow \Halpha (black) and broad
\Halpha, which we use as a point-source reference. Each row shows a different
OB, ordered by increasing image quality. The left column
(panels~\subref{f.spatialha.a}, \subref{f.spatialha.c}
and~\subref{f.spatialha.e}) shows the data, normalised to the peak flux
density, with the blue and red wings of broad \Halpha in blue and red,
respectively (the inset in panel~\subref{f.spatialha.a} displays the
three spectral windows we averaged to obtain the narrow and broad-line
profiles). The bottom panels show the narrow-to-broad ratio, with
increasingly strong departures from unity as image quality improves.
The right column (panels~\subref{f.spatialha.b}, \subref{f.spatialha.d}
and~\subref{f.spatialha.f}) shows the best-fit exponential model of
narrow \Halpha (green), convolved with the PSF (dashed grey),
taken to be the average of the red and blue broad-line wings of \Halpha.
The bottom panels show the $\chi$ residuals, with systematic departures
suggesting a structured narrow-line profile or deviations from a simple
exponential model. The inferred \re is reported in
the top right corner.
}\label{f.spatialha}
\end{figure}

In Fig.~\ref{f.spatialha} we compare the spatial profile of the narrow and broad
\Halpha line as measured in three OBs, ordered top to bottom by increasing image
quality. The left column (panels~\subref{f.spatialha.a}, \subref{f.spatialha.c}
and~\subref{f.spatialha.e}) shows the empirical comparison, illustrating that in
all OBs, narrow \Halpha (black) is more extended than both the blue and red wing
of broad \Halpha (blue and red curves, obtained as the average flux density
within rest-frame 6550--6556~\AA and 6572--6578~\AA). The bottom panels
show the ratio between narrow \Halpha and the broad-\Halpha wings. The strong
and systematic departure from unity (horizontal dashed line) implies that narrow
\Halpha is more extended. Furthermore, the consistent behaviour of the blue and
red curves highlights the stable PSF shape between the blue and red wings of
broad \Halpha, as expected given the short wavelength separation. Crucially,
evidence of spatial extension increases with improving image quality.

In the right column we quantify the spatial extension of \Halpha, adopting as
instrument PSF the average between the spatial profiles of the blue and red
broad-\Halpha wings (grey dashed line). We model the spatial profile of narrow
\Halpha as an exponential, with half-light radius \re. We then convolve this
exponential with the PSF (which already accounts for slit convolution) to
match the data. The best-fit model is shown by the green curve, while the bottom
panels display the fit residuals. From the best-seeing OB (OB03) we infer
$\re = 0.23\pm0.01$~arcsec, corresponding to $440\pm19$~pc at $z=0.1006$. This
value is comparable to the \hst-based \re for the local LRD J1047
\citep[$\re=500$~pc;][]{lin+2026a}, and to the narrow-line sizes inferred in
a lensed LRD at $z=7.04$ \citep{juodzbalis+2026b}. Nevertheless, we also notice
systematic negative residuals near the profile core, indicating that the adopted
model function may not fully capture the galaxy profile. Since the residuals are
mostly negative in the centre and positive at large radii, we speculate that a
S\'ersic index $n>1$ may be preferred over our assumed exponential.

We note that the fit residuals are structured across all three OBs, with the
fiducial exponential model typically over-predicting the flux near the core
($\lesssim1$~arcsec). To test if this reflects the adopted PSF or the source
model, we explored several variations of the model fit. Replacing the
empirical broad-line PSF with a Moffat fit improves the fit only modestly
($\chi^2=14.4$ instead of $19.2$), while symmetrising the empirical PSF alone
does not help. In contrast, replacing the exponential source profile with a
S\'ersic model substantially improves the fit ($\chi^2=7.7$ with the empirical
PSF, and $5.9$ with the Moffat PSF), but the inferred S\'ersic indices are
$n\simeq4.5$ and $n\simeq3.1$, uncharacteristically large for a gas distribution.
Nevertheless, these fits suggest that the dominant mismatch is due to the
adopted source profile rather than to the PSF model. If both the narrow-line
profile and the PSF are symmetrised about their centroids, an exponential model
yields a remarkably smaller $\chi^2=0.4$. In this case, $\re\simeq0.08$~arcsec,
but this approach is very risky, because the source is only marginally resolved;
hence it is hard to assess if the symmetrisation procedure biases the inferred
radius.

\section{Discussion}\label{s.d}

Our spectropolarimetry yields a small number of largely model-independent
results, which we then use to address a sequence of physical
questions. The results are:
\begin{enumerate}
  \item the continuum is polarised at $p_{\rm cont}\simeq1.5$ per cent, while broad \Halpha is less polarised ($p_{\Halpha}=0.58\text{--}0.84$~per cent; Section~\ref{s.r.ss.hapol}) with an offset in the line polarisation angle of $\Delta\vartheta\simeq-48$\textdegree;
  \item the continuum polarisation is nearly grey;
  \item we detect no blue-to-red swing in polarisation angle across the line;
  \item the width $W$ of the exponential model is statistically consistent between polarised and total light (within our large uncertainties);
  \item narrow \Halpha is spatially more extended than the broad-line wings.
\end{enumerate}
The discussion is organised around the physical questions these results help to address. After establishing that the signal is intrinsic and not foreground (Section~\ref{s.d.ss.intrinsic}), we summarise which mechanisms can polarise the source. Given the debated nature of LRDs, we adopt a physics-first approach: we derive our constraints from the data and from general physical arguments, to avoid biasing the interpretation towards familiar source classes. We thus discuss the continuum (Section~\ref{s.d.ss.continuum}), continuum--line differences (Section~\ref{s.d.ss.symmetry}), and the broad \Halpha line (Section~\ref{s.d.ss.linepol}). We defer the comparison with the extensive AGN-polarisation literature to Section~\ref{s.d.ss.analogies}. Finally, Section~\ref{s.d.ss.preferred} provides our preferred interpretation.

\subsection{The polarisation is intrinsic to \texorpdfstring{\target}{J1025}}
\label{s.d.ss.intrinsic}

A purely Galactic foreground dust screen cannot explain the observed spectropolarimetry. Such a screen would impose the same wavelength-dependent Stokes vector on both continuum and line photons, while the data show a localised line effect across broad \Halpha.

At the high Galactic latitude of \target ($b\simeq+54$\textdegree), Galactic foreground dust is minimal: we find $A_{V, \rm MW}=0.14$~mag \citetext{from both \citealp{schlafly+finkbeiner2011} and \textit{Planck}, \citealp{planckdust+2015}; we converted $E(B-V)$ measurements to $A_{V, \rm MW}$ assuming the \citealt{cardelli+1989} law with $R_V=3.1$}. For such low dust columns, the \textit{Planck} 353-GHz polarised-dust maps are noise dominated \citep{planckdust+2015,planckdust+2020}, so the foreground polarisation angle is unconstrained. Nevertheless, even assuming the maximum interstellar polarisation efficiency \citetext{i.e., the empirical \citealt{serkowski+1975} ceiling $p/A_{V, \rm MW}\le3$ per cent mag$^{-1}$, reached only for ideally aligned grains with the magnetic field in the plane of the sky, \citealt{andersson+2015}}, the Galactic foreground along this sightline would be $p_{\rm MW}<0.4$ per cent, below both $p_{\rm cont}$ and $p_{\Halpha}$. Assuming typical efficiencies \citep[2--3 times below the ceiling; e.g.,][]{ fosalba+2002}, the realistic Galactic polarisation is $p_{\rm MW}<0.2$~per cent and thus sub-dominant in our observed signal.
Turning the argument around, if the observed continuum signal were dominated by Galactic extinction, it would require $p_{\rm cont}/A_{V, \rm MW}\simeq10.9$ per cent mag$^{-1}$, several times higher than the maximum efficiency.

We conclude that while Galactic foreground polarisation may still contribute at a sub-dominant level, it cannot dominate the observed signal. Given our inability to accurately measure the foreground polarisation, we quote only the amplitude limit $p_{\rm MW}<0.4$ per cent and do not attempt a Stokes-space foreground subtraction.

The line effect is the key point: even if the foreground contribution were larger than expected for this particular sightline, a smooth foreground screen cannot produce the observed change in polarisation fraction and polarisation angle across the broad line. The polarisation signal must therefore be primarily intrinsic to \target.

\subsection{Polarisation mechanisms and their signatures}
\label{s.d.ss.mechanisms}

Before addressing what polarises the continuum and the broad line, it is useful to review the relevant polarisation mechanisms and their distinguishing signatures. Polarisation can be (i) intrinsic, i.e. generated within the atmosphere of the source itself, through radiation transfer in a scattering-dominated, aspherical, highly ionised atmosphere \citep[in AGN, the accretion-disc atmosphere;][]{laor+1990,coleman+shields1990,kartje+konigl1991,afanasiev+2019}; or extrinsic, driven by (ii) electron scattering in ionised gas outside the source, (iii) dichroic transmission through magnetically aligned dust grains, or (iv) dust scattering. The polarisation amplitudes are set by different quantities.

Intrinsic polarisation is set by the optical depth of the source, by the asphericity of the emitting medium, and by its inclination to the line of sight, reaching, in the optically thick limit, a theoretical maximum of 11.7 per cent for a pure electron-scattering atmosphere viewed edge-on, and decreasing monotonically to $p=0$ with decreasing inclination $i$ \citep{chandrasekhar1960}. Externally scattered light can be polarised at tens of per cent, depending on the scattering angle. Dichroic polarisation is set by the dust column, grain composition, and degree of grain alignment: NIR polarimetry of Seyfert-2 nuclei implies that AGN-local dust is generally inefficient, $dp/dA_V\sim0.1\text{--}0.3$ per cent mag$^{-1}$ \citep[e.g.,][]{young+1995,lumsden+1999}, well below the maximum Galactic value \citep[$\sim3$ per cent mag$^{-1}$, measured at optical wavelengths;][]{serkowski+1975}.

These mechanisms also differ in their wavelength dependence. Electron scattering, whether in the source atmosphere or in external ionised gas, produces grey polarisation. For intrinsic polarisation, deviations from a grey curve can arise from a number of radiative transfer effects \citep{laor+1990,coleman+shields1990,agol+blaes1996,gnedin+2005,afanasiev+2019}, which together predict a non-monotonic wavelength dependence. Uniquely among these mechanisms, electron scattering, when it acts on emission lines and has sufficient optical depth, also produces substantial broadening \citetext{e.g., \citealt{laor2006,chang+2026}, which we discuss in Section~\ref{s.d.ss.broadening}}. Dichroic polarisation peaks near optical wavelengths for Milky-Way-like grains \citep[$\lambda_{\rm max}\simeq0.55~\mum$;][]{serkowski+1975}, shifting with the size of the aligned grains \citep{whittet+1992}, which in AGN may differ from Galactic dust \citep[e.g.,][]{maiolino+2001a}; the cross section for dust scattering, instead, rises towards the blue \citep[cf. the dust- and electron-scattered sightlines in NGC~1068;][]{miller+1991}. Our nearly grey polarisation curve (Fig.~\ref{f.polcol}) is consistent with electron scattering, but our short optical baseline cannot single out a unique mechanism from the polarisation curve alone (Appendix~\ref{a.altmean}).

\subsection{The optical continuum}\label{s.d.ss.continuum}

\subsubsection{A single dominant source}
\label{s.d.ss.continuum.sss.single}

Most LRD models assume a single, dominant source for the optical continuum, be it a massive accreting black hole \citep[e.g.,][]{madau+maiolino2026,rusakov+2026} or a supermassive star \citep[e.g.,][]{nandal+loeb2026}. Here we ask if the data support this assumption.

While our observations partially resolve the host galaxy (or an extended narrow-line region, $\re=440\pm19$~pc; Section~\ref{s.r.ss.spatial}), host-galaxy light cannot explain the continuum in \target, as standard stellar populations are known not to reproduce its observed spectral features \citep{lin+2026a}. Our continuum polarisation also rules out a population of non-standard, emission-line stars. Many distinct and comparable sources would not naturally combine into one coherent continuum Stokes vector, yielding small net polarisation. $N$ comparable sources with polarisation $p^*$ and random polarisation orientations yield an average net polarisation $\approx p^*/\sqrt{N}$; therefore, explaining $p_{\rm cont}$ with $N$ sources requires $p^*\gtrsim p_{\rm cont}\,\sqrt{N}$; comparing the \Halpha luminosity of \target to that of the most luminous emission-line star \citep[NGC~2363~V1, $L_{\Halpha}=10^{38}\,\ergs$;][]{drissen+2001} $N\sim7000$, so the resulting $p^*$ is unphysical. Even for the most luminous LBV stars, estimated at $10^{40\text{--}41}\,\ergs$ \citep{guseva+2024}, $N=7\text{--}70$, yielding $p^*=4\text{--}12$~per cent; while feasible in theory, systematically attaining such large polarisations for random orientations (averaging at 60\textdegree) would exceed the corresponding Chandrasekhar limit of $p^*\simeq2.8$~per cent \citep{chandrasekhar1960}. More decisively, when star clusters are polarised, this is usually explained by external polarisation \citep[e.g.,][]{bijas+2022}. But because this mechanism operates outside of both the continuum and broad lines, it cannot produce the line effect observed across broad \Halpha. Therefore, our data leave one single, dominant source as the most natural explanation for the optical continuum.

We stress that since our spectropolarimetry covers only the rest-frame optical ($\simeq5300\text{--}6650~\AA$), it provides no new constraints on the ongoing debate about the origin of the rest-frame UV continuum: whether it is host- or AGN-dominated \citep[e.g.][]{naidu+2025,degraaff+2025a,sun+2026,deugenio+2025e,torralba+2026b,perez-gonzalez+2026,geris+2026b}. Equally, because we do not cover \OIIIall, we cannot test the compact scenario recently proposed to explain short-timescale variability in an LRD at $z=6.7$ \citep{lambrides+2026}.

\subsubsection{A direct view of the dominant optical continuum source}\label{s.d.ss.continuum.sss.los}

Having established that a single source dominates the optical continuum (Section~\ref{s.d.ss.continuum.sss.single}), we can now ask whether we view this source directly (as in, e.g., type-1 AGN), or only via scattered light (as in, e.g., type-2 AGN).

We first address the extreme case, in which the observed continuum is entirely scattered light: a single hidden continuum source, seen only via a scattering `mirror'. The observed level of polarisation disfavours this scenario: purely reflected light is expected to be highly polarised for generic scattering geometries (Section~\ref{s.d.ss.mechanisms}), well above our observed $p_{\rm cont}\simeq1.5$~per cent. While some Seyfert-2 nuclei with low observed polarisation do exist, their low $p$ is dominated by dilution from unpolarised host starlight within the aperture; when starlight is subtracted, the remaining scattered nuclear light is highly polarised \citep[$p=3\text{--}10$ per cent; e.g.,][]{tran1995}. This dilution channel is impossible in \target, whose optical continuum is not host-galaxy light \citetext{\citealp{ji+2025a,naidu+2025,degraaff+2025a}; for \target itself, see \citealp{lin+2026a,ji+2026a}}, so a reflection-dominated view would require strong geometric cancellation, i.e., a finely tuned, nearly axisymmetric mirror viewed close to face-on.

Even though we disfavour a reflection-only view of the continuum, a composite scenario is still possible, such as a two-component continuum, in which an unpolarised source dominates the total flux, while a second component, hidden from direct view, is seen only in scattered (hence polarised) light. This hypothesis has concrete realisations among proposed LRD models, for example an unpolarised thermal or nebular continuum seen directly, plus an accretion disc seen only through scattering \citep[e.g.,][]{lin+2026a}. Polarimetry disfavours this scenario too, because the polarised flux $p\times I$ has nearly the same spectral shape as the total continuum (Fig.~\ref{f.specpol.f}). If the two hypothesised continuum components had different natures, the hidden component would need to mimic, across our full wavelength range, the spectral shape of the physically different, dominant source -- which would be a remarkable coincidence. Still, we caution that two moderately different continua could still match in shape within our uncertainties, given the short wavelength baseline.

We conclude that the total-light and the polarised-light continua most likely arise from a single dominant component, viewed at least in part directly. This, however does not constrain the nature of the source, nor which mechanism polarises it, which we address next.

\subsubsection{Origin of the optical continuum and its polarisation}
\label{s.d.ss.continuum.sss.contpol}

We now ask what polarises the optical continuum. Having established a single, dominant source (Section~\ref{s.d.ss.continuum.sss.single}), the answer depends on the nature of the continuum source, so we address it separately for three limiting continuum scenarios: nebular dominated, thermal photosphere, and accretion disc.

In a nebular-dominated scenario \citep[e.g.,][]{sneppen+2026b}, where the continuum and broad \Halpha arise from recombination in the same ionised gas, any polarisation mechanism would imprint nearly the same polarisation on both continuum and line. So the observed contrast in both $p$ and $\vartheta$ rules out this scenario for \target. For similar reasons, we can also exclude, for \target, a Compton-thick shell in which the light from a central engine is thermalised by the same gas that produces \Halpha. Variants in which a nebular continuum or the recombination lines are subject to partial reprocessing \citep[e.g.,][]{sneppen+2026b} are not ruled out outright 
in terms of polarised fraction (due to a possible role of non-scattering opacities and RT effects), but remain strongly disfavoured by the polarisation angle (Sections~\ref{s.d.ss.symmetry} and~\ref{s.d.ss.linepol}), clearly indicating that continuum and line arise from different structures, at least in \target.

For a thermal photosphere, intrinsic polarisation is not expected at the observed level. There are two reasons. First, in a cold photosphere \citep[$T\sim 5000$~K;][]{lin+2026a}, the dominant polarisation mechanism is Rayleigh scattering, but we see no evidence of the corresponding steep blue rise in our polarisation curve \citep[e.g., Fig.~\ref{f.polcol.c};][]{auriere+2016}. Second, at least to the extent that blackbody models are assumed to be nearly spherical \citep[][]{begelman+dexter2026,kido+2025,liu+2025}\footnote{The truncated-disc version of the \citealt{liu+2025} model is geometrically closer to an accretion disc compared to the other blackbody models.}, any scattering-induced polarisation undergoes strong geometric cancellation, so the net intrinsic polarisation is expected to be low. On the other hand, non-negligible flattening due to rapid rotation may lead to some polarisation, a quantification of which is beyond the scope of this work (see also Section~\ref{s.d.ss.analogies.sss.stars} below for some known examples of polarised thermal photospheres). External electron- or dust-scattering in a compact region near the source is in principle a natural polariser; thus, even a perfectly spherical photosphere can acquire the observed $p_{\rm cont}$ this way. The expected rise of the polarised fraction towards the blue is not observed, mildly disfavouring dust, although the short baseline cannot conclusively exclude it (Section~\ref{s.d.ss.mechanisms} and Appendix~\ref{a.altmean}). Since the external scatterer also intercepts the broad line, invoking this mechanism for the continuum must carefully account for the polarisation properties of the broad line too (Section~\ref{s.d.ss.linepol}). Dichroic dust polarisation could also explain the observations. However, thermalised models are often invoked to explain the red optical continuum of LRDs as a low-temperature effect, as opposed to dust reddening \citep{kido+2025,liu+2025}. Therefore, invoking substantial dichroic-dust polarisation would reintroduce the dust column that this scenario is designed to avoid. Quantitatively, to achieve the observed $p_{\rm cont}=1.5$ per cent, dichroic polarisation would require $A_V>0.5$~mag \citep[under the maximum efficiency of][]{serkowski+1975}, and possibly much larger columns for the lower efficiencies commonly inferred for AGN dust (Section~\ref{s.d.ss.mechanisms}), disfavouring this mechanism.

Finally, a hot, optically thick accretion disc \citetext{possibly attenuated by dust or dense gas, \citealp{inayoshi+maiolino2025,ji+2025a,naidu+2025,torralba+2026b,pacucci+ferrara2026}} is subject to intrinsic polarisation due to electron scattering within its own atmosphere (Section~\ref{s.d.ss.mechanisms}). Because a disc is flattened, per-cent-level polarisation follows from its geometry alone, increasing with inclination, so our measured $p_{\rm cont}\simeq1.5$ per cent is most naturally accommodated by an inclined viewing angle. For a semi-infinite, pure-scattering atmosphere, the maximum polarised fraction is $p=11.7$ per cent for $i=90$\textdegree, decreasing to $p\simeq1.5$ per cent for an inclination angle $i\simeq45$\textdegree \citep{chandrasekhar1960}, although this value should be viewed as a lower limit, since multiple physical effects can act to decrease intrinsic polarisation already at $i=90$\textdegree (Section~\ref{s.d.ss.mechanisms}). We note that even an inclination much higher than $i=45$\textdegree does not conflict with the direct view established in Section~\ref{s.d.ss.continuum.sss.los}, because the relatively modest dust columns inferred for LRDs \citetext{ranging from $A_V =0~mag$ for the models of \citet{naidu+2025,torralba+2026b}, to $A_V\simeq0.5 \text{--}5$~mag for the models of \citealp{madau+maiolino2026,brazzini+2026,ji+2026b,pacucci+ferrara2026}, as opposed to $A_V\simeq20\text{--} 40$~mag in type-2 AGN; e.g., \citealp{young+1995,lumsden+1999}} leave the intrinsic optical continuum from the accretion disc unattenuated, or mildly attenuated but still visible \citep{madau+maiolino2026}. While intrinsic polarisation is natural for this model, external electron- or dust-scattering contributions are also viable mechanisms, subject to the same caveats as for the thermalised photosphere model. Dichroic polarisation is possible as well, and -- unlike for thermalised models -- accretion discs can accommodate (and in fact often invoke) dust along the line of sight. Although, in the context of an accretion disc model, dust should generally be regarded as a possible additional contributor, rather than the main agent, of continuum polarisation.

In summary, the total-light and polarised-light continua likely arise from the same, single dominant component (Sections~\ref{s.d.ss.continuum.sss.single} and~\ref{s.d.ss.continuum.sss.los}). Among the scenarios, nebular continuum alone is ruled out for \target; a nearly spherical thermal photosphere with $T\simeq5000$~K seems also disfavoured for lack of the characteristic Rayleigh-scattering signature at relatively blue wavelengths. While viable for all scenarios, external scattering is subject to additional constraints from the broad-line properties, which we discuss next. This leaves an intrinsically polarised accretion disc as the simplest framework.

\subsection{The angle offset requires broken axial symmetry}
\label{s.d.ss.symmetry}

One conclusion is independent of which polarisation mechanisms operate: whatever polarises the line or the continuum, an axisymmetric configuration can only produce polarisation angles of 0\textdegree or 90\textdegree with respect to the projection of the symmetry axis in the plane of the sky, so the observed intermediate $\vartheta$ offset (Section~\ref{s.r.ss.hapol} and Table~\ref{t.measurements}) requires broken axial symmetry \citep[e.g.,][]{goosmann+gaskell2007}. The basic argument is very general. In a coordinate system ($Q^\prime$, $U^\prime$) aligned with the symmetry axis of the system, the assumed axial symmetry implies a symmetry by reflection in the plane defined by the symmetry axis and the line of sight, which prohibits values of (wavelength-integrated) $U^\prime$ different from 0 and therefore polarisation angles different from 0\textdegree or 90\textdegree.

For instance: in the case of internal electron scattering, the expected polarised fraction depends on both the flattening and the optical depth of the emitting surface, as well as the line of sight to the observer \citep{angel1969}. A gradual change in optical depth corresponds to a gradual change in $Q^\prime$ from positive to negative values, while $U^\prime$ is always zero for symmetry reasons. As a result, local maxima of the polarised fraction are achieved in the optically thin and thick limits, with a shift of 90\textdegree in polarisation angle between the two, while zero polarisation is in general achieved at some intermediate optical depth, whose value depends on the axis ratio and viewing angle. Crucially, for an axisymmetric system, there is no value of the optical depth and no line of sight towards which the emitted light can have a wavelength-integrated polarisation angle different from 0\textdegree or 90\textdegree with respect to the projected symmetry axis. Therefore, if both the continuum and the BLR are separately polarised by internal scattering in an axisymmetric structure, we must conclude that the two structures are strongly misaligned (by $\sim40\text{--}50$\textdegree) with respect to one another.

Similarly, external scattering (by either ionised gas or dust) can produce
intermediate polarisation angles only if the scatterer does not share the same
symmetry axis as the source, while dichroic transmission requires the magnetic
field to be misaligned.

As a first but very general conclusion, we can exclude, at least for \target, a
dust-free homogeneous spherical configuration, as well as any combination of nested
aligned axisymmetric structures without dust. Removing any one or more of these
assumptions opens the door to possible solutions.

We identify three general symmetry-breaking mechanisms, which we will use in 
the following: a line-emitting or line-scattering structure inclined with respect
to the axis of the continuum polariser; large-scale clumpiness or irregularity
of the line-emitting or scattering medium; and dichroic transmission through an
external dust screen whose grain-aligning magnetic field is misaligned with the
continuum polarisation axis.

\subsection{BLR geometry and polarisation}\label{s.d.ss.linepol}

Different models have been proposed for the geometry and nature of the BLR in LRDs, based on several partially interconnected questions: (i) is the BLR a rotating disc, or is it more similar to a quasi-spherical structure? (ii) What physical processes are responsible for the broadening of the lines? (iii) Is the BLR physically related to the dense gas absorbers detected in Balmer emission lines? (iv) Is the BLR related to the mechanism responsible for the reddening of the optical continuum in LRDs compared to ordinary AGN? And (v) how to explain the similarities and differences between LRDs and LBDs\footnote{Acronym for `Little Blue Dots', with some similarities with LRDs, but no Balmer break and overall bluer SEDs, see e.g.\ \cite{brazzini+2026}}?

Here below we discuss what our polarisation measurements tell us about the
first (geometrical) question. The second, partially related question about the broadening mechanism is discussed in Section~\ref{s.d.ss.broadening}. The other questions are touched upon when relevant, although addressing all of them in a complete and exhaustive way is beyond the scope of this work.

\subsubsection{Quasi-spherical BLR}\label{s.d.ss.linepol.sss.sphere}

A perfectly spherical, homogeneous BLR cannot be intrinsically polarised, from symmetry. To be consistent with our data, a model of this kind requires an external polariser (electron or dust scattering, or dichroic transmission), which must also be misaligned with respect to the polarisation axis of the continuum. Similar to what discussed for the continuum (Section~\ref{s.d.ss.continuum.sss.contpol}), achieving the observed $p_{\Halpha}=0.58\text{--}0.84$ by dichroic transmission requires a dust attenuation of at least $A_V\simeq0.2\text{--}0.4$~mag, assuming the maximum theoretical efficiency of 3 per cent mag$^{-1}$ (Section~\ref{s.d.ss.mechanisms}), or significantly higher levels ($A_V \simeq 2\text{--}5$~mag) if a typical AGN efficiency of 0.1-0.3 per cent mag$^{-1}$ (see Section~\ref{s.d.ss.mechanisms}) is assumed. Note that, in the current literature, spherical BLR configurations are generally proposed together with the assumption of dust-free screens, and the high observed Balmer decrements are driven primarily by collisional excitation and/or radiative transfer effects \citep[e.g.,][]{chang+2026,matthee+2026}. So while possible, the dust solution does not match any of the currently proposed models to our knowledge.

On the other hand, a quasi-spherical BLR does not need to be exactly spherical, but may have some flattening. Also, recent quasi-spherical BLR models of LRDs tend to have a non-homogeneous structure, with thick inflows on the equator and thin outflows on the poles \citep{sneppen+2026a,matthee+2026}. This may be related to the LRDs/LBDs phenomenology \citep{matthee+2026}. Models of this kind have a preferred direction; therefore, in addition to external polarisation mechanisms (as above), they may also have intrinsic polarisation due to internal scattering. Note that internal scattering may also be related to the BLR
broadening mechanism (Section~\ref{s.d.ss.broadening.sss.escatt}). The exact polarised fraction then depends on many factors (flattening, latitudinal structure, optical depth, line of sight), so a quantitative assessment would require a dedicated theoretical study and is left for future work. A robust conclusion for \target is that, if intrinsic polarisation of a quasi-spherical BLR dominates, then the preferred axis of the BLR must have a strong (40--50\textdegree) misalignment with respect to the symmetry axis of the continuum.

\subsubsection{Rotating disc}\label{s.d.ss.linepol.sss.disc}

We now discuss the implications of our polarisation measurements for a scenario in which the BLR of \target is a rotating disc. We keep the discussion general, but we note at the start that a rotating disc is invoked, in particular, by the unification model proposed by \citet{madau+maiolino2026}. Like the \citet{matthee+2026} model, the \citet{madau+maiolino2026} model also addresses the LRD/LBD dichotomy, but this time invoking dust (rather than just dense gas) being concentrated along the equator, more similar to the classical unification model of ordinary AGNs \citep{antonucci1993,urry+padovani1995}. Further similarities and differences with the classical AGN unification model are discussed in Section~\ref{s.d.ss.analogies}, while implications for the line broadening mechanism are left for Section~\ref{s.d.ss.broadening.sss.virial}. Here we focus on polarisation and geometry.

Because a rotating disc is highly flattened, intrinsic polarisation is possible, and actually likely. If this is the dominant mechanism, however, the observed line-to-continuum offset in polarisation angle requires the BLR disc to be strongly tilted ($\sim40\text{--}50$\textdegree) compared to the accretion disc, just as discussed in Section~\ref{s.d.ss.linepol.sss.sphere}.

Scattering by an external ionised medium is also possible. However, satisfying the broken symmetry requires that the scatterer be out of axis. We should also in general expect that the same ionised gas will scatter and polarise light from the accretion disc as well, in which case it must be in an opportune 3D position to intercept the BLR and the continuum with vastly different angles, as observed. In particular, the distance of the scatterer from the central source cannot be much larger than that of the BLR, otherwise both BLR and continuum would be seen under a similar scattering angle.

Dust scattering is an appealing option for a rotating disc, because it can break the axial symmetry \emph{locally}, i.e. at each wavelength: this produces a blue-to-red angle swing for pure rotation \citep{smith+2005}, or an M-shaped angle structure for rotation plus a wind \citep{lira+2021}. Neither, however, produces a global, velocity-integrated offset like that in \target unless the \emph{global} axial symmetry is also broken --- by a dust plane or wind misaligned with the disc, or by large-scale clumpiness that samples the scattering geometry incompletely \citep{smith+2005}. In all cases, dust scattering requires a line of sight that does not pass through the dust screen; it is therefore disfavoured for \target within the \citet{madau+maiolino2026} model, in which a red LRD such as \target is seen precisely \emph{through} the screen.

Dichroic transmission through a dust screen with misaligned magnetic field is not only possible, but in fact natural: within the \citet{madau+maiolino2026} framework, a red object like \target needs a dust screen as an intrinsic model component, rather than an \textit{ad-hoc} addition. From the observed $p_{\Halpha}$, we derive the same dust attenuation as for the spherical BLR models, but this time $A_V=2\text{--}5$~mag aligns very well with the model's requirements. As a
consistency check, taking the broad line to be intrinsically unpolarised and to carry only the screen term, $p_{\rm d}\simeq0.58$~per cent at
$\vartheta_{\rm d}\simeq-34$\textdegree, and subtracting this term in Stokes space from the observed continuum polarisation, the intrinsic continuum polarisation is $\simeq1.7$~per cent at $\vartheta\simeq+22$\textdegree: thus the screen slightly rotates and slightly reduces the intrinsic disc polarisation, without cancelling it -- all within the reasonable polarisation efficiency budget of $dp/dA_V\simeq0.1\text{--}0.3$~per cent~mag$^{-1}$.
Finally, the misalignment between the two polarisers arises naturally, because the grain-aligning magnetic field does not need to share the orientation of the nuclear disc axis \citep{balbus+hawley1991}: the broken axial symmetry is then expected, rather than fine-tuned. We recall that the presence of significant dust extinction in some LRDs has been challenged by the lack of corresponding emission in the MIR \citep[e.g.,][]{setton+2025}, but this assumes isotropic dust coverage and may not apply if extinction occurs only along specific lines of sight.

\subsubsection{Irregular or chaotic BLR}\label{s.d.ss.linepol.sss.irregular}

Finally, the BLR itself might be highly irregular or chaotic. The observed signal in this case could arise from a structure lacking any special symmetry and dominated by one or more large clumps with a random orientation with respect to the accretion disc and our line of sight. A scenario of this kind could easily explain a large angle offset, regardless of the exact polarisation (or broadening) mechanism. On the other hand, it may not be easy, within this framework, to explain the absence of a similarly strong velocity offset of the BLR emission from the systemic velocity, or the low observed levels of variability \citetext{\citealp{zhang+2026a,burke+2026,liu+2026}, but see \citealp{ji+2025a,furtak+2025a,lambrides+2026} for a different view}. We refrain from further speculation in this regard, as to our knowledge models of this kind are so far absent from the LRD literature, but may be worth some investigation in the light of our measurement for \target. A larger sample of LRDs with spectropolarimetry could certainly also be helpful in this respect. Within a chaotic model, the key prediction is that different objects should have a wide variety of angle offsets, with a random, or close to random, distribution.

\subsection{Line broadening mechanism}\label{s.d.ss.broadening}

We now turn from what polarises the broad line to what broadens it, discussing two contrasting limits: microscopic broadening by electron scattering \citep{rusakov+2026}, and kinematic broadening by virial motions \citep[e.g.,][]{juodzbalis+2026b}. Beyond the physical understanding of LRDs \citep[and other exponential-wing AGN;][]{brazzini+2026}, the distinction bears directly on the inferred mass of the central black hole and its accretion rate \citep{maiolino+2024,rusakov+2026}, with a profound impact on black-hole seeding and growth.

 It is also useful to recall that non-Gaussian line profiles (including exponential wings) are not unique to LRDs, but are seen in blue AGN \citep{brazzini+2026} and in luminous quasars \citetext{e.g., \citealp{nagao+2006,kollatschny+zetzl2013}, Trefoloni, in~prep.}. Both virial motions and electron scattering have been discussed, prior to the discovery of LRDs, in the literature on line broadening for ordinary AGN (e.g.\ \citealt{laor2006}).

\subsubsection{Electron-scattering broadening}\label{s.d.ss.broadening.sss.escatt}

In the electron-scattering scenario, \Halpha is broadened by a scattering medium either co-spatial with the \Halpha production site \citep{torralba+2026b}, or interposed between the BLR and the observer at least along our line of sight \citep{rusakov+2026}.

In the recent literature, electron-scattering broadening is often invoked by quasi-spherical BLR models, although in principle it could apply to disc or irregular morphologies as well. Regardless of the exact geometry, for small optical depths $\tau_\mathrm{e} \lesssim 1$ this mechanism predicts that $p\times I$ be broader than $I$, because photons in the exponential wings have been scattered and hence polarised, while those in the unscattered core should not be polarised. This effect is seen, for example, in NGC~1068, where broad \Halpha viewed through electron-scattering regions is substantially broader than the same line viewed through dust-scattering ones \citep{miller+1991}. Although promising in principle, the discriminatory power of this test is unfortunately quite limited in \target. This is because, within the scattering-broadening model, a prominent exponential wing (as in \target) is only possible as a result of a relatively large $\tau_\mathrm{e}$ ($\gtrsim1$), which in turn implies a large suppression ($e^{-\tau_\mathrm{e}}$) of the unpolarised core signal. Further difficulties arise from contamination of the line core by absorption features, and by the narrow-line region and host galaxy, which may or may not be polarised by an independent mechanism. Quantitatively, our data are consistent with the width of polarised and total light being the same (Section~\ref{s.r.ss.polflux}), as well as with a toy scattering model (Section~\ref{s.r.ss.toyfit}), with residuals indicative of complex structures near line core. We cannot therefore reach a clear conclusion on this matter based on our data, but we note that further progress could be made by targeting with high SNR objects with a larger core-to-wing ratio, no prominent absorption features and no evidence of a spatially resolved NLR.

We recall that our toy electron-scattering fit (Section~\ref{s.r.ss.toyfit}) requires $\tau_\mathrm{e}\sim1$ (constrained by the transmitted-core fraction) and $p\sim0.8$~per cent (constrained by the polarised flux). In principle, this combination of parameters could be used to put constraints on the required flattening and viewing angle of a hypothetical ellipsoidal cocoon \citep[e.g.,][]{angel1969}. This is left for future investigation. Importantly, we recall that the inferred optical depth is a consequence of fitting exponential wings with a scattering kernel, which, as we discuss below, is not the only possible approach to the problem.

\subsubsection{Virial broadening}\label{s.d.ss.broadening.sss.virial}

Electron scattering is not the only mechanism able to produce the exponential line shape of LRDs: recent stratified-BLR models reproduce the same exponential wings as LRDs by purely kinematic broadening, without invoking electron scattering as the primary mechanism \citep{scholtz+2026b,madau+2026}.

In principle, virial broadening is possible in any geometry, including a quasi-spherical configuration, provided that the opacity to electron scattering $\tau_\mathrm{e}$ is sufficiently small. It is therefore not possible to formulate a completely general prediction of the spectropolarimetric signature of virial broadening. With this caveat in mind, we restrict our attention to the LRD/LBD unification model by \cite{madau+maiolino2026}, which invokes virial broadening specifically in the context of a rotating disc (see Section~\ref{s.d.ss.linepol.sss.disc}). A rotating disc is also among the most widespread virialised models for standard AGN. It is directly observed in some high-luminosity quasars \citep{gravity+2018,gravity+2024,gravity+2026}, and inferred in standard AGN from the blue-to-red swing in $\vartheta$ across the broad lines \citep[e.g.,][]{smith+2005}; this includes some low-luminosity narrow-line Seyfert-1 galaxies, a luminosity regime closer to LRDs like \target \citep[e.g.,][]{afanasiev+2019}.

Within the \citet{madau+maiolino2026} model, the favoured BLR polarisation mechanism for \target is dichroic transmission (Section~\ref{s.d.ss.linepol}), whose efficiency is achromatic across the line and therefore implies identical $p\times I$ and $I$ widths. This is consistent with our results, although, as mentioned in Section~\ref{s.d.ss.broadening.sss.escatt}, we cannot consider this as a decisive proof in favour of one particular broadening mechanism.

In addition to targeting other objects with a larger core-to-wing ratio, the \citet{madau+maiolino2026} model could also be tested by targeting LBDs, for which the favourite polarisation mechanism would be equatorial dust scattering and the distinctive feature would be a blue-to-red swing in polarisation angle, without a systematic (angle-integrated) offset. A detected swing would be transformative, as it would both confirm the origin of the line broadening and open an independent avenue for measuring the black-hole mass \citep{afanasiev+2015}.

\subsection{Phenomenological analogies}
\label{s.d.ss.analogies}

Having discussed the physical constraints, and before presenting a synthesis (Section~\ref{s.d.ss.preferred}), we now step back to ask where comparable polarisation signatures have been observed. On their own, these phenomenological analogies are not conclusive; still, they show that neither the continuum-to-line depolarisation nor the large angle offset are unique to \target -- even though no single object class reproduces the full combination.

\subsubsection{Type-1 AGN spectropolarimetry}
\label{s.d.ss.analogies.sss.agn}

Since LRDs have broad Balmer lines, the closest phenomenological comparison is with type-1 AGN spectropolarimetry. In the Seyfert sequence discussed by \citet{smith+2002,smith+2004,smith+2005}, equatorial-scattering Seyfert~1 nuclei can show a blue-to-red swing in polarisation angle across broad lines, while polar-scattered Seyfert~1 nuclei \citep{smith+2004} show stronger continuum polarisation and Seyfert-2 scattering signatures. Near face-on systems can have low net polarisation because of azimuthal cancellation.

While \target does not fully match any one of these limiting cases, type-1 AGN show remarkable object-to-object diversity \citep{smith+2002,smith+2004}: broad-line depolarisation, core dips, and wing polarisation are all common in type-1 AGN, so \target would not be unique, in this respect. The closest analogue of \target in polarisation is Mrk~279, which displays both depolarised \Halpha relative to the continuum, and different polarisation angle between continuum and \Halpha (evaluated without continuum subtraction), $\Delta\vartheta \simeq 61$\textdegree \citep{smith+2002}. However, the total-light picture is completely different, as Mrk~279 is X-ray bright \citep{constantini+2007} and time variable \citep{bachev+strigachev2003}.

Especially interesting analogues are narrow-line Seyfert-1 galaxies (NLSy1), which can show broad-line depolarisation and complex line-profile behaviour \citep{goodrich1989,smith+2002}. The comparison with NLSy1 is also interesting for other reasons. Due to their relatively narrow broad lines, it is easier to identify the signatures of electron-scattering broadening \citep{laor2006}. Relatively narrow and forbidden \FeIIperm emission was first identified in 1~Zw\,\textsc{i} \citep{veron-cetty2004}, and recently discovered in LRDs at both low and high redshifts \citetext{e.g., \citealt{lin+2026a,ji+2026a} at $z\lesssim 0.2$; \citealt{tripodi+2025, deugenio+2025e,torralba+2026b} at $z=5\text{--}7$}. Finally, the NLSy1 Akn-564 is the nearest nitrogen-loud AGN \citep{yang+2025}, echoing the high nitrogen abundance in some LRDs \citep{ji+2025b} and LBDs \citep{ji+2024}. Still, \citet{smith+2002} argue that NLSy1s do not form a single polarimetric family, and that their \Halpha polarisation properties are indistinguishable from those of ordinary Seyfert-1 galaxies. In our case, we can take the strong object-to-object differences within the same AGN sub-classes as a warning that a single observation is unlikely to be representative of the entire LRD class.

In the unification model of \citet{madau+maiolino2026}, LRDs are the obscured counterparts of LBDs, consistent with the dichroic-screen picture for \target (Section~\ref{s.d.ss.linepol.sss.disc}). The polarisation nonetheless differs from classical type-2 (Seyfert-2) nuclei, in which the broad lines, seen only in scattered light, are at least as polarised as the continuum \citep{antonucci+miller1985,smith+2002,tran2003}, unlike the depolarised broad \Halpha in \target; this is natural if \target is instead viewed through modest, dichroic obscuration rather than through equatorial or polar scattering, unlike in classical type-2 AGN.

A final useful AGN comparison is with the quasars studied by \citet{kishimoto+2004, kishimoto+2005,kishimoto+2008}, in which the polarised flux is dominated by the continuum and the broad lines are strongly suppressed. In objects like Ton~202 \citep{kishimoto+2003}, the detection of a `Balmer-edge' feature in polarised light echoes the observations of LRDs in direct light \citep{labbe+2024,setton+2025b}.

\subsubsection{Emission-line stars}\label{s.d.ss.analogies.sss.stars}

Emission-line stars offer an instructive phenomenological analogy, though we stress that the physical drivers, energy source, and scales differ entirely from those of \target (as we quantify at the end of this section).

In several classes of line-emitting stars, such as Herbig Ae, Be\textsuperscript{\textcolor{white}{*}} and B[e] stars, as well as luminous blue variables (LBVs), the continuum is intrinsically polarised due to a flattened geometry, interpreted as an accretion or decretion disc (for Herbig stars, \citealt{vink2002}) or the stellar surface itself (for LBVs, \citealt{davies+2005}). \Halpha is often found to be polarised differently from the continuum, including several examples of depolarised lines, especially among Herbig Be stars (\citealt{vink2002}) and LBVs (\citealt{davies+2005}), similar to the depolarisation pattern we observe in \target. Line depolarisation in Herbig Be stars has been interpreted as a signature of the line forming over a less flattened, more symmetric volume compared to the continuum \citep{oudmaijer+drew1999,vink2002}. For some of the Herbig Be stars and LBVs with depolarised \Halpha, there is also a measurable angle offset, although it is usually modest ($\lesssim$ 10 deg).\footnote{We note, however, that larger offsets are sometimes found in emission-line stars where \Halpha is over-polarised, or has an irregular polarisation pattern, relative to the continuum (\citealt{vink2002}).} For Herbig Be stars, \citet{oudmaijer+drew1999} attribute the \Halpha polarisation angle offset to superimposed interstellar and circumstellar dust polarisation over an otherwise unpolarised line, which matches our dichroic scenario. In the LBVs of \citet{davies+2005}, line polarisation is instead considered to be intrinsic and is thought to trace wind asphericity and clumpiness. Furthermore, the polarisation angle in these objects is found to be variable with time, over a timescale of decades (\citealt{davies+2005}), reinforcing an interpretation of asphericity and clumpiness in their stellar winds. This erratic behaviour is not unexpected given the violent, spatially complex mass-loss phenomena known to occur in these systems, resonating with the `chaotic BLR' scenario briefly sketched in Section~\ref{s.d.ss.linepol.sss.irregular} (see also \citealt{naidu+2026} for a differently flavoured analogy between LBVs and LRDs). Overall, there are interesting similarities between the polarisation properties of emission-line stars and \target, although we have found no example that matches exactly the pattern we observe in our target. 

Besides these similarities, it should be recalled that these classes of stars are a disparate set of systems, ranging from pre-main sequence objects, to late-stage intermediate-mass stars, to very massive stars on the main-sequence. The structural analogy is however suggestive: the fact that such a diverse set of objects can still present a large number of common spectral properties suggests that the underlying physical structure and processes are not unique, and to the contrary must be relatively easy to attain. One such model is the classic B[e]-star picture of \citet{zickgraf+1985,zickgraf+1986}: a fast polar wind pinched by a slow, dense equatorial flow, the stellar analogue of an outflow pinched by an accretion disc, reminiscent of the inflow/outflow model proposed for LRDs \citep{sneppen+2026a,matthee+2026}. In Zickgraf's model, the disc generates high auroral-to-nebular line ratios by collisional excitation in very dense, low-ionization gas (e.g., auroral \SIIall[4069,4071], \NIIL[5755]), also observed in LRDs \citep{ji+2026a,deugenio+2025e}. These parallels motivate emission-line stars as a phenomenological point of comparison (not a physical model) for some of the processes at play in LRDs (Appendix~\ref{a.herbig}).

Nevertheless, the luminosity, physical scale, and power source of \target are those of an accretion-powered source, not of a nuclear-powered one. As discussed in Section~\ref{s.d.ss.continuum.sss.single}, \target is 7000 times more luminous in \Halpha than the most luminous emission-line star \citep{drissen+2001}, and 7--70 times more luminous than candidate LBV stars in metal-poor starburst dwarf galaxies \citep{guseva+2024}. For high-luminosity LRDs \citep{wang+2024,juodzbalis+2024, labbe+2024}, the flux mismatch rises to several orders of magnitude.

\textcolor{white}{$^{*}$Named after their super-solar beryllium abundance (Herbig, priv.\ comm.).}

\subsection{A preferred picture for \texorpdfstring{\target}{J1025} and an outlook}
\label{s.d.ss.preferred}

Our preferred (but not unique) interpretation is the simplest one surviving the constraints of Sections~\ref{s.d.ss.continuum.sss.contpol} and~\ref{s.d.ss.linepol}: the continuum is not nebular dominated, and is polarised intrinsically, as expected for an accretion disc; most broad-\Halpha emission arises over a larger region, which is either inclined with respect to the continuum polariser or seen through a misaligned dichroic dust screen. This naturally explains why the continuum is more polarised than broad \Halpha, and why the broad line has a weak but distinct polarisation angle.

If framed within the unification scheme by \cite{madau+maiolino2026}, the easiest solution for the symmetry break is dichroic polarisation of the BLR by a magnetised dust screen, with a misaligned magnetic field. In this picture, the key difference with respect to standard type-1 AGN is a modest dust obscuration ($A_V\simeq2\text{--}5$~mag). In particular, in this interpretation, \target is geometrically type-2-like -- obscured and reddened -- but with a much weaker obscuring torus compared to classical type-2 AGN ($A_V\gtrsim20$~mag). The much smaller dust column does not hide the broad lines nor the accretion-disc continuum, so the type-1 signatures, and features such as the Balmer break, appear in direct light rather than only in polarised light.

More in general, our findings are consistent with a broader LRD/LBD picture in which a central compact nuclear structure manifests itself as an LRD or an LBD along different lines of sight, as a consequence of orientation effects and in particular different columns of either dense gas \citep{matthee+2026} or dust \citep{madau+maiolino2026}, or both.

Our polarimetry data do not rule out a quasi-spherical cocoon, but symmetry breaking means that the model must be more complex than a nearly spherical shell, so that the BLR must have a preferred axis (by either flattening or latitudinal stratification of gas column density) which differs markedly from the symmetry axis of the continuum. Alternatively, an aligned or even spherical cocoon could also be reconciled with the data by invoking dichroic polarisation through a magnetised dust screen, with $A_V\sim 2\text{--}5$~mag (Section \ref{s.d.ss.linepol.sss.sphere}).

A highly irregular geometry of the BLR, characterised by e.g. chaotic accretion, large-scale granularity and no special symmetry could also be a route to explain our data, although it may be in contrast with other constraints (Section~\ref{s.d.ss.linepol.sss.irregular}) and certainly requires a statistical sample to be further investigated.

As for what broadens the line, both electron scattering and virial motions are consistent with our data: the exponential wing widths of the polarised and total flux are statistically indistinguishable (Section~\ref{s.d.ss.broadening}).

A larger sample would help to better constrain the polarisation mechanisms in LRDs, as well as LBDs. For the scenario of dichroic transmission, more data would allow to test the prediction that BLR polarisation without angle swings should occur mainly along lines of sight typical of LRDs and not along those of LBDs \citep{madau+maiolino2026,brazzini+2026}. At the same time, the dust scattering mechanism from a rotating BLR could instead be revealed by the measurement of the angle-swing signature in relatively face-on LBDs (Section~\ref{s.d.ss.broadening.sss.virial}). Finally, more measurements of differential widths of polarised and total-intensity lines for objects with different core-to-total fraction would help to put stringent constraints on electron-scattering as a viable broadening mechanism (Section~\ref{s.d.ss.broadening.sss.escatt}).

\section{Conclusions}\label{s.conc}

We have presented VLT/FORS2 optical linear spectropolarimetry of the local Little Red Dot \target \citetext{$z\simeq0.1$; `Lord of LRDs', \citealp{ji+2026a}; `The Egg', \citealp{lin+2026a}}, spanning rest-frame $\sim5300\text{--}6650~\AA$. 

\begin{itemize}

\item The continuum near \Halpha is polarised at $p_{\rm cont}\simeq1.5$\,per cent, with polarisation angle $\vartheta_{\rm cont}\simeq+12$\textdegree. $p_{\rm cont}$ is comparable to some intrinsically polarised Seyfert-1 nuclei. A purely Galactic foreground origin is excluded, so the polarisation is intrinsic to \target and points to a single, dominant continuum source rather than to many stellar sources.

\item Broad \Halpha is also polarised, but less than the continuum ($p_{\Halpha}=0.58\text{--}0.84$\,per cent). Pure dilution by an intrinsically unpolarised line does not fully explain the data: the continuum-subtracted line shows a significant polarisation-angle offset from the continuum, $\Delta\vartheta\simeq-48$\textdegree\ (line PA $\vartheta_{\Halpha}=-34\pm3$\textdegree). Continuum and broad-line are produced in different locations and further require a break of axial symmetry in the system.

\item The continuum polarisation is only weakly chromatic ($dp/d\lambda\simeq-0.26\pm0.15$~per cent per rest-frame $1000~\AA$), so the present optical data do not single out intrinsic disc polarisation, dust, or grey electron scattering as the dominant continuum-polarising channel.

\item The \Halpha profile in polarised light is not detectably broader than the profile in total light: modelling the narrow polarised core, the exponential wing width of the polarised flux is $W_{p\times I}=490\pm130~\kms$, statistically consistent with $W_{I}=550\pm10~\kms$ from the total-light fit, but the large errorbars on $W_{p\times I}$ make the width test unconstraining.

\item We derive a robust geometric constraint: the strong line angle offset requires broken axial symmetry; suitable symmetry breaking mechanisms are a broad-line emitting or scattering region misaligned relative to the continuum, an asymmetric external scattering mirror, or dichroic dust transmission, with the grain-aligning magnetic field misaligned relative to the continuum polarisation axis.

\end{itemize}

Taken together, the simplest surviving interpretation (Section~\ref{s.d.ss.preferred}) is a continuum intrinsically polarised in a compact, flattened region, with broad \Halpha forming over a larger volume and polarised along a different axis; the nebular-dominated continuum is ruled out, and any line-scattering medium must be inhomogeneous, or misaligned with respect to the symmetry axis of the continuum source. Both virial motions and electron scattering are consistent with the data; deeper spectropolarimetry of \target and a sample of local LRDs can rule out one or the other scenario.

Future spectropolarimetric observations on larger samples will also be crucial to determine whether the symmetry break we identified in \target is unique to this object or a general property of the LRD (and possibly LBD) population. If an asymmetry is found to be common, a systematic study of the relation (or lack thereof) of the line-continuum polarisation angle offset with other properties (e.g., broad-band spectrum, line shape) could shed light on the geometry and ultimate nature of this mysterious class of objects.

\section*{Acknowledgements}

Based on observations made with ESO Telescopes at the La Silla Paranal Observatory under programme ID 115.29FY.
We are grateful to the ESO Director General, Xavier Barcons, for awarding Director's Discretionary Time for this programme, and to Dr Henri M. J. Boffin and the Paranal observatory staff for their support in executing the observations.
FDE, RM, XJ, JS, IJ, and RGP acknowledge support by the Science and Technology Facilities Council (STFC), by
the ERC through Advanced Grant 695671 ``QUENCH'', and by the UKRI Frontier Research grant RISEandFALL.
RM also acknowledges funding from a research professorship from the Royal Society.
AM acknowledges the MIRACLE INAF 2024 GO grant ``A JWST/MIRI MIRACLE: Mid-IR Activity of Circumnuclear Line Emission'', the INAF Large Grant 2022 ``The metal circle: a new sharp view of the baryon cycle up to Cosmic Dawn with the latest generation IFU facilities'', the grant PRIN-MUR 2020ACSP5K\_002 financed by European Union -- Next Generation EU, and project PRIN-MUR project ``PROMETEUS'' financed by the European Union -- Next Generation EU, Mission 4 Component 1 CUP B53D2300475000.
CRA acknowledges support from the Agencia Estatal de Investigaci\'on of the Ministerio de Ciencia, Innovaci\'on y Universidades (MCIU/AEI) under the grant ``Tracking active galactic nuclei feedback from parsec to kiloparsec scales'', with reference PID2022$-$141105NB$-$I00 and the European Regional Development Fund (ERDF). 
XL and XF acknowledge support from the NSF grant AST-2308258.
MB acknowledges support from the Next Generation EU funds within the National Recovery and Resilience Plan (PNRR), Mission 4 - Education and Research, Component 2 - From Research to Business (M4C2), Investment Line 3.1 - Strengthening and creation of Research Infrastructures, Project IR0000034 – ``STILES - Strengthening the Italian Leadership in ELT and SKA''.
ZC acknowledges support from the National Key R\&D Program of China (grant no. 2023YFA1605600), the National Natural Science Foundation of China (\#12525303), the Tsinghua University Initiative Scientific Research Program, and the New Cornerstone Science Foundation through the XPLORER PRIZE.
SC acknowledges support by European Union's HE ERC Starting Grant No. 101040227 -- WINGS.
IJ also acknowledges support by the Huo Family Foundation through a P.C. Ho PhD Studentship.
ST acknowledges support by the Royal Society Research Grant G125142.
FS acknowledges support from NSF award AST-2513040.

This work made extensive use of the freely available \href{http://www.debian.org}{Debian GNU/Linux} operating system.
We used the \href{http://www.python.org}{Python} programming language \citep{vanrossum1995}, maintained and distributed by the Python Software Foundation. We made direct use of Python packages
{\sc \href{https://pypi.org/project/astropy/}{astropy}} \citep{astropy+2013,astropy+2018},
{\sc \href{https://pypi.org/project/corner/}{corner}} \citep{foreman-mackey2016},
{\sc \href{https://pypi.org/project/emcee/}{emcee}} \citep{foreman-mackey+2013},
{\sc \href{https://pypi.org/project/jwst/}{jwst}} \citep{alvesdeoliveira+2018},
{\sc \href{https://pypi.org/project/matplotlib/}{matplotlib}} \citep{hunter2007},
{\sc \href{https://pypi.org/project/numpy/}{numpy}} \citep{harris+2020},
{\sc \href{https://pypi.org/project/astro-prospector/}{prospector}} \citep{johnson+2021} \href{https://github.com/bd-j/prospector}{v2.0},
{\sc \href{https://pypi.org/project/PyNeb/}{pyneb}} \citep{luridiana+2015},
{\sc \href{https://pypi.org/project/python-fsps/}{python-fsps}} \citep{johnson_pyfsps_2023},
{\sc \href{https://pypi.org/project/pysersic/}{pysersic}} \citep{pasha+miller2023},
{\sc \href{https://github.com/honzascholtz/qubespec/}{qubespec}} \citep{scholtz+2025},
and {\sc \href{https://pypi.org/project/scipy/}{scipy}} \citep{jones+2001}.
We also used the software packages {\sc \href{https://github.com/cconroy20/fsps}{fsps}} \citep{conroy+2009,conroy_gunn_2010}, {\sc \href{https://www.star.bris.ac.uk/~mbt/topcat/}{topcat}} \citep{taylor2005}, {\sc \href{https://github.com/ryanhausen/fitsmap}{fitsmap}} \citep{hausen+robertson2022} and {\sc \href{https://sites.google.com/cfa.harvard.edu/saoimageds9}{ds9}} \citep{joye+mandel2003}.

\section*{Data Availability}

The raw data will be available on the \href{https://archive.eso.org/eso/eso_archive_main.html}{ESO Science Archive Facility}, under programme ID 115.29FY.001. The reduced and combined spectra used in this work are publicly available on Zenodo, at \href{https://doi.org/10.5281/zenodo.6926671}{https://doi.org/10.5281/zenodo.6926671}. The reduced SDSS spectra used for the calibration are available from the SDSS website, at \href{http://dr13.sdss.org/sas/dr13/sdss/spectro/redux/26/spectra/lite/1747/spec-1747-53075-0337.fits}{this link.} This work has made use of the BeSS database, operated at LESIA, Observatoire de Meudon, France: \href{http://basebe.obspm.fr}{http://basebe.obspm.fr}. 

\bibliographystyle{mnras}
\bibliography{spollord}

\appendix

\section{Off-target null spectra}\label{app:nullspectra}

As an independent check on the Stokes-flux noise, we extracted off-target
``null'' spectra from the same reduced FORS2 two-dimensional products used for
the science spectrum. For each of the eight retained OBs, we placed two
apertures above and below the target trace, offset by 24 spatial pixels from the
ordinary and extraordinary target positions. Each aperture had the same
half-width of 8 pixels used for the corresponding beam extraction. We then
formed Stokes-flux spectra directly from the ordinary--extraordinary beam
differences, producing 16 null realisations in total.

\begin{figure}
  \centering
  \includegraphics[width=\columnwidth]{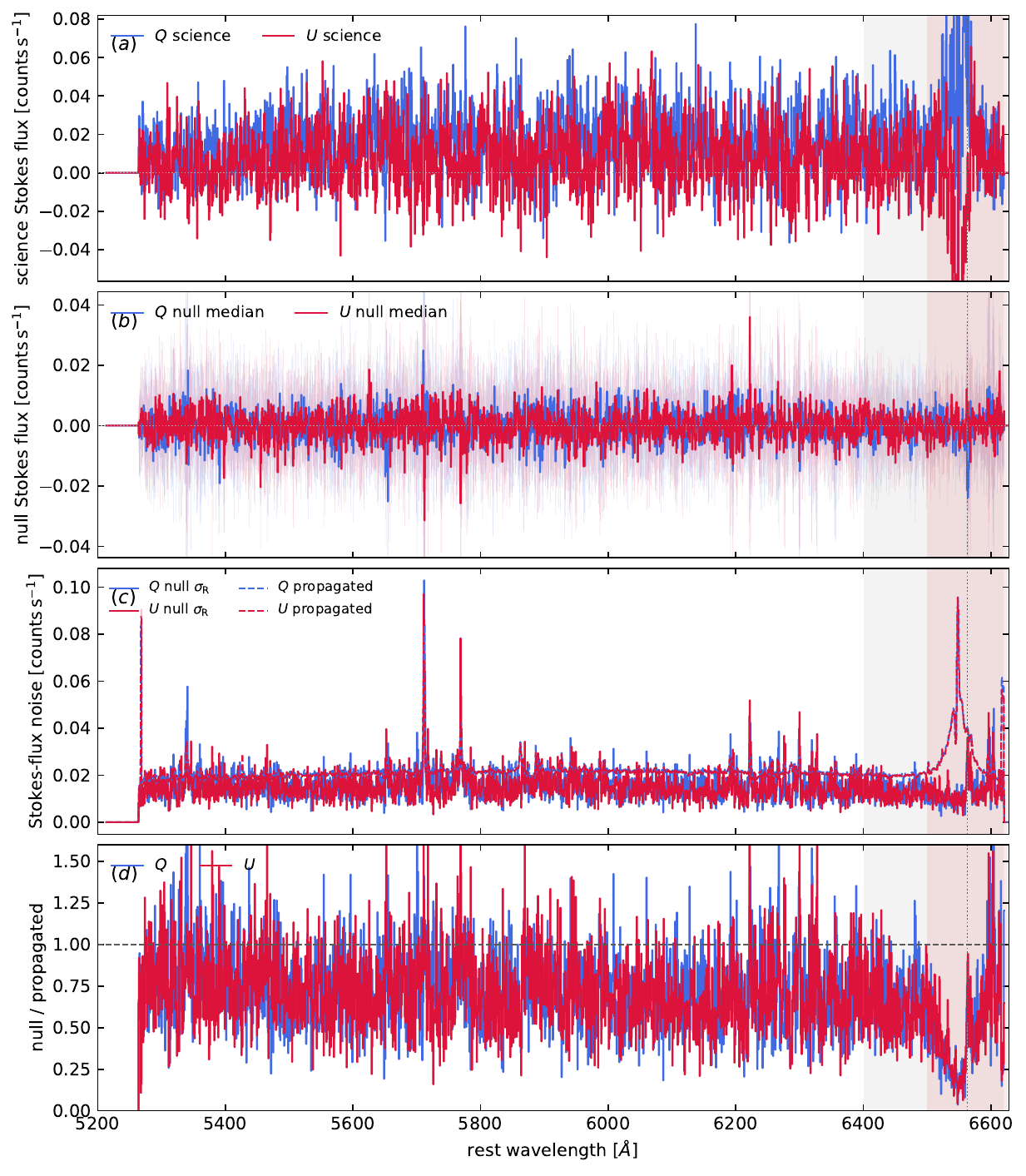}
  \caption{Off-target null-spectrum check in Stokes flux. Panel (a) shows the
  median science $Q$ and $U$ spectra. Panel (b) shows the median off-target
  null spectra, with shaded bands marking the robust scatter of the 16 null
  realisations. Panel (c) compares the robust null scatter with the propagated
  Stokes-flux errors used for the science spectrum. Panel (d) shows their
  ratio; values below unity indicate that the off-source empirical noise is
  smaller than the propagated error. The grey band marks the blue wing of
  broad \Halpha, while the red band is the main line.
  }\label{f.nullspectra}
\end{figure}

Fig.~\ref{f.nullspectra} compares the null spectra with the science Stokes-flux
products. The median null spectra remain close to zero across the wavelength
range, including the \Halpha region. As reference noise level, we use the
robust null scatter $\sigma_{\rm R}(\lambda)$, defined in each wavelength bin
as 1.4826 times the median absolute deviation of the 16 null realisations. This
procedure estimates their standard deviation while remaining insensitive to
outliers. In the continuum window
5200--6400~\AA, the median absolute null residuals are only 0.22 times the
robust null scatter $\sigma_{\rm R}$ in both $Q$ and $U$. Across the \Halpha window
6400--6620~\AA, the corresponding values are 0.23 in both Stokes components,
and in the line-core window 6500--6625~\AA they are 0.23 and 0.25 for $Q$ and
$U$, respectively. We therefore find no evidence for a coherent off-source
Stokes residual that could mimic the line signal.

The empirical null scatter is also smaller than the propagated Stokes-flux
errors used in the science analysis: for both $Q$ and $U$, the median ratios
between the null scatter and the propagated error are 0.69 in the continuum
window, and 0.57 across broad \Halpha. The mismatch is by construction, since
the null apertures contain no photon noise. Indeed, the null-to-propagated
noise ratio decreases where the science intensity rises (Spearman $\rho=-0.33$
between $I/I_{\rm cont}$ and the mean $Q,U$ ratio).

\section{Alternative averaging schemes}\label{a.altmean}

The polarisation values we report are sensitive to the averaging method and weighting scheme employed. Here we discuss the implications for our analysis. Contrary to expectations, inverse-variance weighting (IVW) gives noisier averages. This was inferred by taking $u$ and $q$ and by comparing their standard deviation across line-free spectral regions between the IVW average and the standard unweighted average. 

In addition, we also employ the median instead of the mean when performing adaptive binning. The mean is nominally more appropriate than the median for linear quantities such as $U$ and $Q$. However, when binning the data with fixed bin size, we noticed increased noise compared to the median-averaged bins, hence we resorted to using the median everywhere, i.e. also for the adaptively binned data. When estimating the uncertainties, we upscale the variance by the canonical value of $\text{\textpi}/2$, to account for the larger uncertainties of the median relative to the mean.

Crucially, we verified that neither using IVW instead of unweighted, nor employing the mean instead of the median affects our main results, with the difference being $\lesssim 0.1$~per cent, below our fiducial systematics floor of $\simeq 0.4$~per cent (set by the unknown foreground polarisation, Section~\ref{s.data}). We tested that any differences are not due to outliers, since \textsigma-clipping the data has negligible effect on the stacks and binning.

Regardless of the method, we find that broad \Halpha is still less polarised than the continuum and with a different angle, and the polarised-flux spectrum still peaks at \Halpha. There is, however, a difference in the polarisation colour. The continuum-colour fit is somewhat steeper than for the fiducial median stack, with $dp/d\lambda=-0.41\pm0.12$ per cent per 1000~\AA\ (3.5~\textsigma), driven primarily by the steeper $dq/d\lambda=-0.35\pm0.12$ per cent per 1000~\AA. The resulting index difference is $\Delta\beta=-1.31\pm0.42$ (3.1~\textsigma). We thus warn that the grey polarisation spectrum is not a foregone conclusion. 

\begin{figure}
  \centering
  \includegraphics[width=\columnwidth]{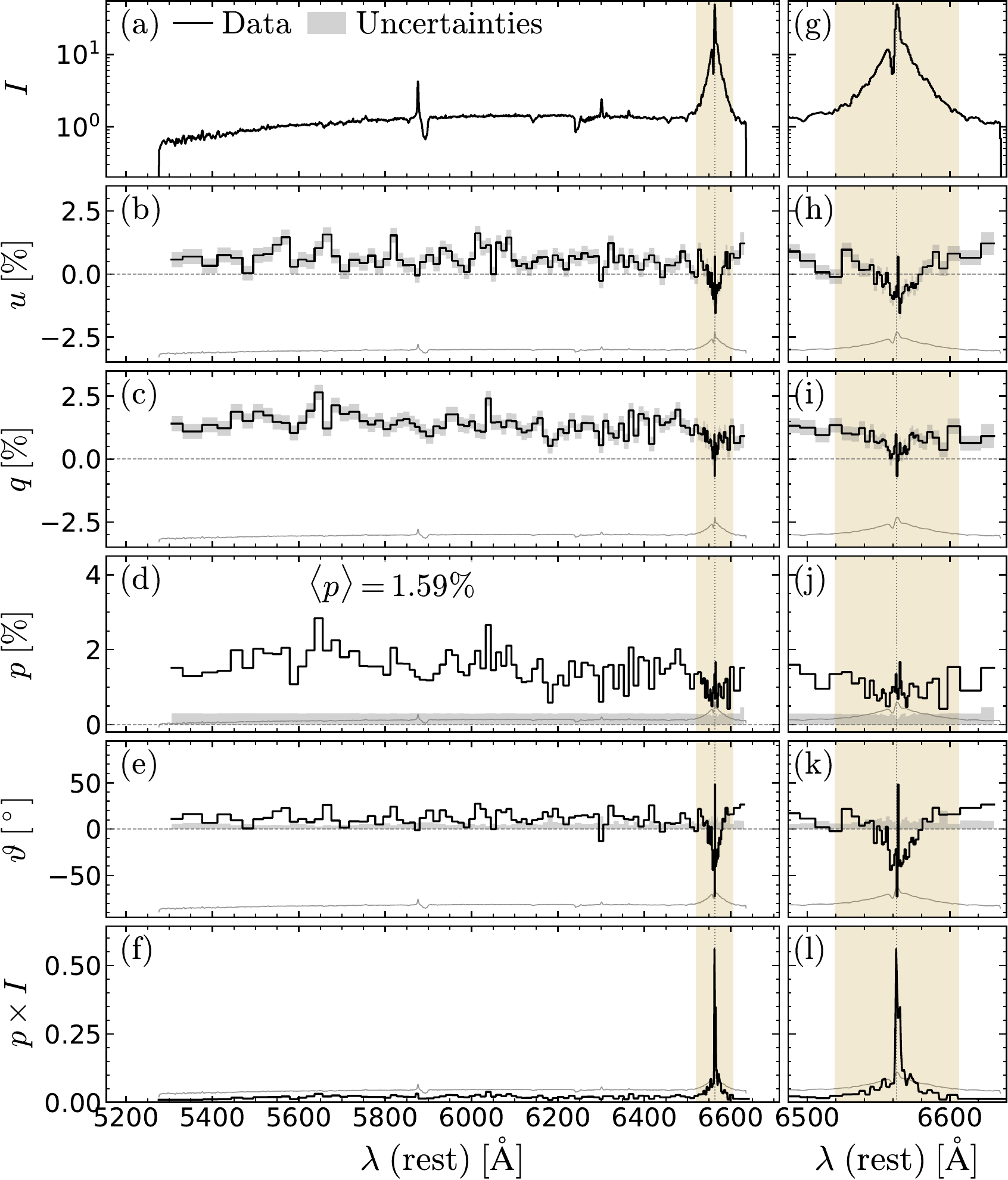}
  \caption{Adaptive-binning spectropolarimetry using a clipped-mean stack, shown as a robustness comparison to Fig.~\ref{f.specpol}. The qualitative behaviour is unchanged, but the polarisation colour becomes bluer, as described in the text.
  }\label{f.specpolclip}
\end{figure}

\section{Flux-only electron-scattering model}\label{a.notoyes}

\begin{figure}
  {\phantomsubcaption\label{f.notoyes.a}
   \phantomsubcaption\label{f.notoyes.b}}
  \centering
  \includegraphics[width=\columnwidth]{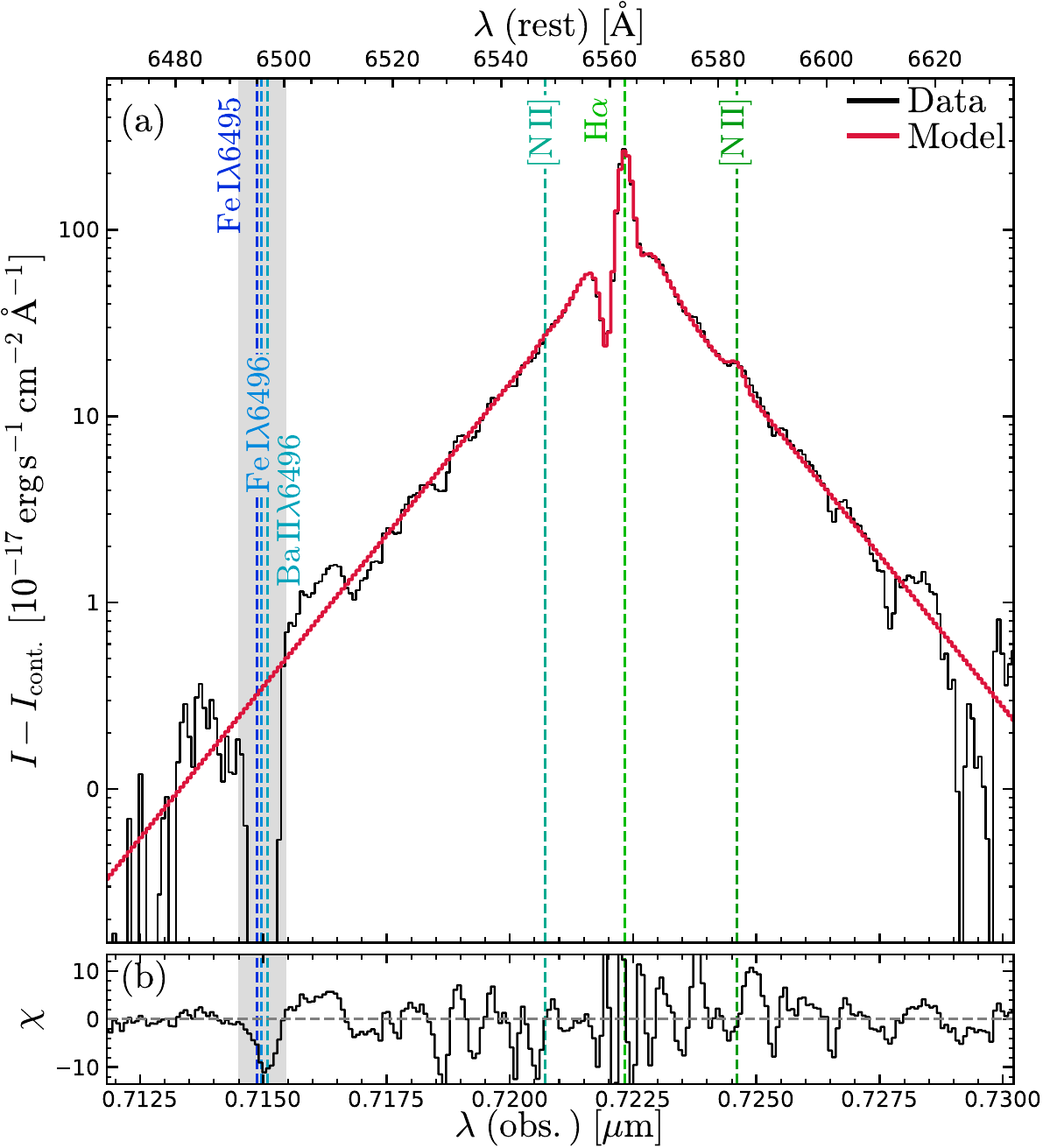}
  \caption{Same as Fig.~\ref{f.toyes}, but fitting only Stokes $I$, without the
    polarisation data. The similarity of the residual structure to that in
    Fig.~\ref{f.toyes} shows that the dominant model mismatch is already present
    in the flux spectrum itself.}\label{f.notoyes}
\end{figure}

Here we show the best-fit model obtained by removing polarisation from the model described in Section~\ref{s.r.ss.toyfit} and therefore fitting the model only to the total-intensity spectrum. The best-fit spectrum is rendered in Fig.~\ref{f.notoyes.a}, which we find indistinguishable from the spectrum of Fig.~\ref{f.toyes.a}. Crucially, the fit residuals (Fig.~\ref{f.notoyes.b}) show the same high-significance undulations we reported in Fig.~\ref{f.toyes.b}, demonstrating that the data--model mismatch reported in Section~\ref{s.r.ss.toyfit} is not driven by the inclusion of $p$ in the toy model, but is intrinsic to the spectral flux density, which is the only quantity we model here.

\section{The curious case of star \lrdstar}\label{a.herbig}

In addition to the eponymous emission lines, Herbig Be and B[e] stars also display other characteristics in common with LRDs, such that it is instructive to study them. Of course, these stars can scarcely be considered a homogeneous family, bar the defining presence of a blue continuum and prominent emission lines. For example, the archetypal Herbig Be star \textgamma\ Cas has low-EW \Halpha ($\simeq-25~\AA$) with a narrow line profile dominated by rotation \citep[equatorial velocity $v_\mathrm{eq}\simeq 450~\kms$;][]{archer+2025}. To add to the mismatch, \textgamma\ Cas is also a strong source of thermalised X-rays (possibly arising from the interaction between the stellar atmosphere and the surrounding decretion disc).

\begin{figure}
  {\phantomsubcaption\label{f.lrdstar.a}
   \phantomsubcaption\label{f.lrdstar.b}}
  \centering
  \includegraphics[width=\columnwidth]{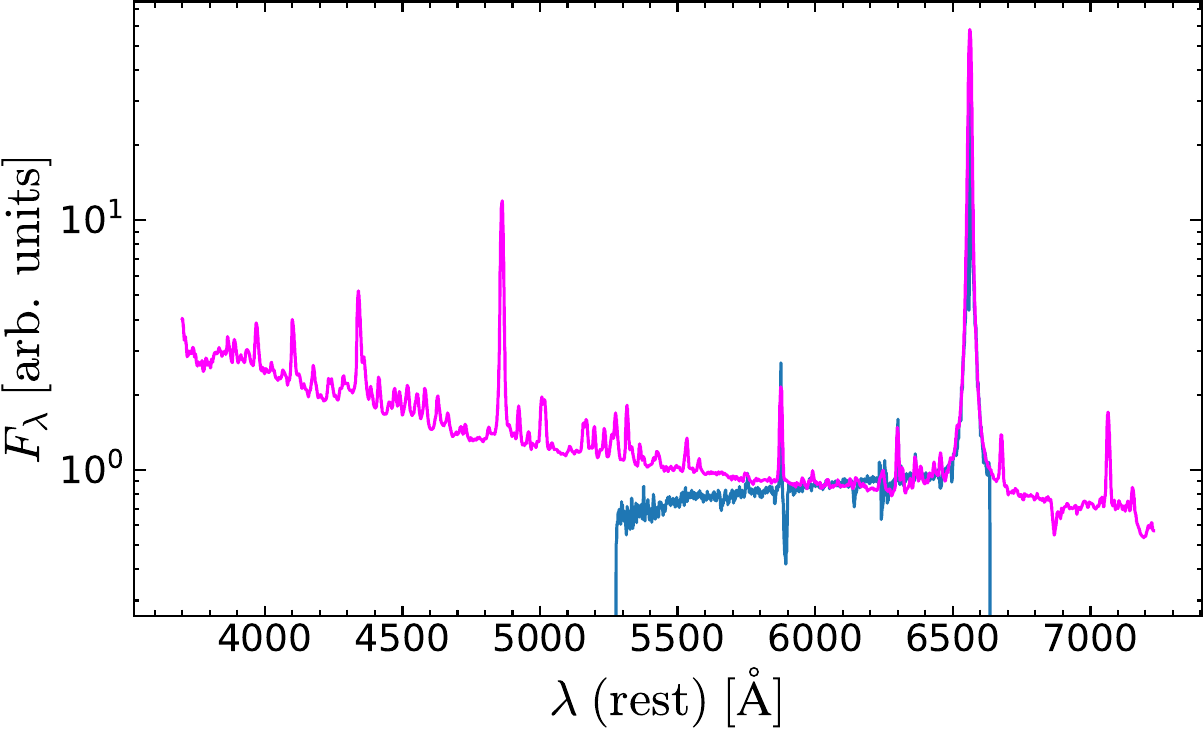}
  \caption{Comparison of \target to \lrdstar; there are many similarities and differences, but some of the similarities are in rather peculiar properties, such as auroral-line emission and Balmer-line properties.
  }\label{f.lrdstar}
\end{figure}

However, in some members of this peculiar class of stars, the spectral similarities with LRDs are much stronger. The B[e] star \lrdstar (Fig.~\ref{f.lrdstar}) is a case in point. This compact, post-AGB planetary-nebula B[e] star \citetext{cPNB[e]; \citealt{lamers+1998}} is thought to be undergoing a rapid transition phase towards a true PN. It shows prominent \Halpha with $EW(\Halpha)\simeq-900~\AA$ \citep{jaschek+1996}, higher than many LRDs \citep[cf. $-340~\AA$ for \target;][]{ji+2026a}. The continuum-subtracted line profile displays the same exponential-like wings as in \target (Fig.~\ref{f.lrdstar.b}), and the Balmer decrement is $F(\Halpha)/F(\Hbeta)\simeq6$, despite the remarkably blue continuum colour. As for other emission lines, \lrdstar displays extremely high auroral-to-strong emission-line ratios (\SIIall[4069,4071] vs \SIIall, \NIIL[5755] vs \NIIL, \OIIIL[4363] vs \OIIIL) also seen in some well-studied LRDs at both low and high redshifts \citetext{e.g., \citealt{ji+2026a} and \citealt{deugenio+2025e}}, which some authors consider evidence for a two-zone density structure \citep{deugenio+2025e}. The forest of permitted \FeIIperm and forbidden \FeII emission is also worth mentioning \citetext{cf. \citealt{jaschek+1996} and \citealt{lin+2026a, tripodi+2025,ji+2026a,torralba+2026b,deugenio+2025e,lambrides+2025}}. 

Another suggestive parallel between LRDs and B[e] stars is represented by the classic Zickgraf B[e]-star picture \citep{zickgraf+1985,zickgraf+1986}: a fast, highly ionized polar wind plus a slow, dense equatorial outflow. According to \citet{zickgraf+1985}, this model can produce most of the peculiar spectral features seen in B[e] stars and, therefore, in LRDs (Fig.~\ref{f.lrdstar}). Translated to an AGN, the polar component would provide an ionized scattering channel for the continuum and broad lines, while the denser equatorial flow would dominate low-ionization line formation and obscuration, giving rise, e.g., to metal and Balmer absorption. In such a geometry the continuum can be polarised mainly by the polar component, yet broad \Halpha can be only weakly polarised if most of the line photons are produced in a larger region than the disc. The Zickgraf-style two-component outflow provides a useful phenomenological template for an LRD in which a fast ionized polar medium and a slower dense equatorial medium coexist, echoing the models of \citet{sneppen+2026a} and \citet{matthee+2026}.

\label{lastpage}
\end{document}